\documentclass[conference]{IEEEtran}
\IEEEoverridecommandlockouts
\usepackage{cite}
\usepackage{amsmath,amssymb,amsfonts}
\usepackage{graphicx}
\usepackage{textcomp}
\usepackage{xcolor}
\usepackage[ruled,vlined,noend]{algorithm2e}
\usepackage[noend]{algpseudocode}
\usepackage{mathtools}
\usepackage{float}
\usepackage[super]{nth}
\usepackage{enumitem}
\usepackage{balance}
\usepackage[normalem]{ulem}
\usepackage{tabularx}
\usepackage{subcaption}
\usepackage{multirow}
\usepackage[cal=dutchcal]{mathalfa}

\DeclarePairedDelimiter{\ceil}{\lceil}{\rceil}
\DeclarePairedDelimiter\floor{\lfloor}{\rfloor}

\def\BibTeX{{\rm B\kern-.05em{\sc i\kern-.025em b}\kern-.08em
    T\kern-.1667em\lower.7ex\hbox{E}\kern-.125emX}}
\begin{document}
\newcounter{example}[section]
\renewcommand{\theexample}{\nthesection.\arabic{example}}
\newenvironment{example}{
     \refstepcounter{example}
     {\vspace{1ex} \noindent\bf  Example  \theexample:}}{
     \vspace{1ex}} 


\newcounter{definition}[section]
\renewcommand{\thedefinition}{\nthesection.\arabic{definition}}
\newenvironment{definition}{
     \refstepcounter{definition}
     {\vspace{1ex} \noindent\bf  Definition  \thedefinition:}}{
     } 

\newcounter{theorem}[section]
\renewcommand{\thetheorem}{\nthesection.\arabic{theorem}}
\newenvironment{theorem}{\begin{em}
        \refstepcounter{theorem}
        {\vspace{1ex} \noindent\bf  Theorem  \thetheorem:}}{
        \end{em}} 

\newcounter{lemma}[section]
\renewcommand{\thelemma}{\nthesection.\arabic{lemma}}
\newenvironment{lemma}{\begin{em}
        \refstepcounter{lemma}
        {\vspace{1ex}\noindent\bf Lemma \thelemma:}}{
        \end{em}} 

\newcommand{\proofsketch}{\noindent{\bf Proof Sketch: }}
\newcommand{\myproof}{\noindent{\emph{Proof:} }}

\newcommand{\nthesection}{\arabic{section}}

\newcommand{\eop}{\hspace*{\fill}\mbox{$\Box$}}

\newcommand{\stitle}[1]{\vspace{0.5ex} \noindent{{\bf #1}}}

\newcommand{\sstitle}[1]{\vspace{0.5ex} \noindent{\textit{ #1}}}
\newcommand{\ssstitle}[1]{\vspace{1ex} \noindent{\textbf{ #1}}}

\newcommand{\kw}[1]{{\ensuremath {\mathsf{#1}}}\xspace}
\newcommand{\bkw}[1]{{\ensuremath {\mathsf{\textbf{#1}}}}\xspace}

\newcommand{\kwnospace}[1]{{\ensuremath {\mathsf{#1}}}}
\newcommand{\ltt}{\kw{LTT}}
\newcommand{\arr}{\kw{arrive}}
\newcommand{\vp}{\kw{p}}

\newcommand{\bellf}{{\sc Bellman-Ford}\xspace}
\newcommand{\bfalgo}{{\sc OR}\xspace}

\newcommand{\aalgo}{{\sc KDXZ}\xspace}

\newcommand{\dijk}{{\sc Dijkstra}\xspace}
\newcommand{\dalgo}{{\sc Two-Step-LTT}\xspace}

\newcommand{\dtalgo}{{\sc DOT}\xspace}

\newcommand{\genfunc}{{\sl timeRefinement}\xspace}
\newcommand{\pathc}{{\sl pathSelection}\xspace}
\newcommand{\fifo}{{\sl FIFO}\xspace}

\newcommand{\st}{starting time\xspace}
\newcommand{\sti}{starting-time interval\xspace}
\newcommand{\stsi}{starting-time subinterval\xspace}
\newcommand{\stsis}{starting-time subintervals\xspace}
\newcommand{\stis}{starting-time intervals\xspace}
\newcommand{\ti}{time interval\xspace}
\newcommand{\tis}{time intervals\xspace}
\newcommand{\ttime}{travel time\xspace}
\newcommand{\ttimea}{travel time\xspace}
\newcommand{\at}{arrival time\xspace}
\newcommand{\ata}{arrival-time\xspace}
\newcommand{\atf}{arrival-time function\xspace}
\newcommand{\ati}{arrival-time interval\xspace}
\newcommand{\ats}{arrival times\xspace}
\newcommand{\ed}{edge delay\xspace}
\newcommand{\eda}{edge-delay\xspace}
\newcommand{\edf}{edge-delay function\xspace}
\newcommand{\eds}{edge delays\xspace}
\newcommand{\wt}{waiting time\xspace}

\newcommand{\g}{\overline{g}}
\newcommand{\iend}{\tau}

\newcommand{\argmin}{\operatornamewithlimits{argmin}}

\newcommand{\myhead}[1]{\vspace{.05in} \noindent {\bf #1.}~~}
\newcommand{\cond}[1]{(\emph{#1})~}
\newcommand{\op}[1]{(\emph{#1})~}

\newcommand{\qc}{\ensuremath{Q^c}}

\newcommand{\rewrite}{\kw{XPathToReg}}

\newcommand{\upparen}[1]{\ensuremath{\mathrm{(}}{#1}\ensuremath{\mathrm{)}}}
\newcommand{\func}[2]{\funcname{#1}\upparen{\ensuremath{#2}}}
\newcommand{\funcname}[1]{\ensuremath{\mathit{#1}}}

\newcommand\AS{\textbf{as}\ }
\newcommand{\xsltsize}{\small}

\newcommand{\X}{{\cal X}}
\newcommand{\sem}[1]{[\![#1]\!]}
\newcommand{\NN}[2]{#1\sem{#2}}
\newcommand{\pcdata}{{\tt str}\xspace}

\newcommand{\exa}[2]{{\tt\begin{tabbing}\hspace{#1}\=\hspace{#1}\=\+\kill #2\end{tabbing}}}
\newcommand{\ra}{\rightarrow}
\newcommand{\la}{\leftarrow}
\newcommand{\rsa}{\_} 
\newcommand{\Ed}[2]{E_{{\scriptsize \mbox{#1} \rsa \mbox{#2}}}}
\newenvironment{bi}{\begin{itemize}
        \setlength{\topsep}{0.5ex}\setlength{\itemsep}{0ex}\vspace{-0.6ex}}
        {\end{itemize}\vspace{-1ex}}
\newenvironment{be}{\begin{enumerate}
        \setlength{\topsep}{0.5ex}\setlength{\itemsep}{0ex}\vspace{-0.6ex}}
        {\end{itemize}\vspace{-1ex}}
\newcommand{\ei}{\end{itemize}}
\newcommand{\ee}{\end{enumerate}}

\newcommand{\mat}[2]{{\begin{tabbing}\hspace{#1}\=\+\kill #2\end{tabbing}}}
\newcommand{\m}{\hspace{0.05in}}
\newcommand{\ls}{\hspace{0.1in}}
\newcommand{\beqn}{\begin{eqnarray*}}
\newcommand{\eeqn}{\end{eqnarray*}}

\newcounter{ccc}
\newcommand{\bcc}{\setcounter{ccc}{1}\theccc.}
\newcommand{\icc}{\addtocounter{ccc}{1}\theccc.}

\newcommand{\oneurl}[1]{\texttt{#1}}
\newcommand{\tabstrut}{\rule{0pt}{4pt}\vspace{-0.1in}}
\newcommand{\tabstruct}{\rule{0pt}{8pt}\\[-2ex]}
\newcommand{\stab}{\rule{0pt}{8pt}\\[-2.2ex]}
\newcommand{\sstab}{\rule{0pt}{8pt}\\[-2.2ex]}

\newcommand{\eat}[1]{}

\newfloat{tcm}{thp}{loa}
\floatname{tcm}{Recursive \sql}

\def\subfigcapskip{2pt}


\newcommand{\rdms}{{\sc rdbms}\xspace}
\newcommand{\sql}{{\sc sql}\xspace}
\newcommand{\dbms}{{\sc dbms}\xspace}

\newcommand{\cfig}{Fig.~}
\newcommand{\ctab}{Table~}
\newcommand{\csec}{Section~}
\newcommand{\cdef}{Definition~}
\newcommand{\cthm}{Theorem~}
\newcommand{\clem}{Lemma~}
\newcommand{\cequ}[1]{Equation~(#1)}
\newcommand{\SG}{\mathbf{SG}}
\newcommand{\SA}{\mathbf{SA}}
\renewcommand{\AA}{\mathbf{AA}}

\newcommand{\xml}{{\sl XML}\xspace}
\newcommand{\xlink}{{\sl XLink}\xspace}
\newcommand{\xpath}{{\sl XPath}\xspace}
\newcommand{\xpointer}{{\sl XPointer}\xspace}
\newcommand{\rdf}{{\sl RDF}\xspace}
\newcommand{\tc}{{\sl TC}\xspace}
\newcommand{\dfs}{{\sl DFS}\xspace}
\newcommand{\DAG}{{\sl DAG}\xspace}
\newcommand{\DAGs}{{\sl DAG}s\xspace}
\newcommand{\grail}{{\sl GRAIL}\xspace}
\newcommand{\yesgrail}{{\sl Yes-GRAIL}\xspace}
\newcommand{\code}{\kw{code}}
\newcommand{\sit}{\kw{sit}}
\newcommand{\psit}{{\cal P}_{sit}}
\newcommand{\yescode}{{\sl Yes-Label}\xspace}
\newcommand{\nocode}{{\sl No-Label}\xspace}
\newcommand{\entry}{\kw{entry}\xspace}
\newcommand{\yngindex}{{\sl YNG-Index}\xspace}
\newcommand{\rqrun}{{\sl RQ-Run}\xspace}
\newcommand{\citeseerx}{{\sl citeseerx}\xspace}
\newcommand{\gouniprot}{{\sl go-uniprot}\xspace}
\newcommand{\uniprot}{{\sl uniprot150}\xspace}

\long\def\comment#1{}

\newcommand{\scc}{strongly connected component\xspace}
\newcommand{\sccs}{strongly connected components\xspace}
\newcommand{\sscc}{\kw{SCC}}
\newcommand{\ssccs}{\kwnospace{SCC}s\xspace}
\newcommand{\sccg}{\kwnospace{SCC}\textrm{-}\kw{Graph}}
\newcommand{\strongc}{\leftrightarrow}
\newcommand{\nstrongc}{\nleftrightarrow}
\newcommand{\emscc}{\kwnospace{EM}\textrm{-}\kw{SCC}}
\newcommand{\dfsscc}{\kwnospace{DFS}\textrm{-}\kw{SCC}}
\newcommand{\dfstree}{\kwnospace{DFS}\textrm{-}\kw{Tree}}
\newcommand{\len}{\kw{len}}
\newcommand{\dep}{\kw{depth}}
\newcommand{\tdep}{\kw{drank}}
\newcommand{\tlink}{\kw{dlink}}
\newcommand{\vedges}{up-edges\xspace}
\newcommand{\vedge}{up-edge\xspace}
\newcommand{\cvedge}{Up-Edge\xspace}

\newcommand{\drsscc}{\kwnospace{1P}\textrm{-}\kw{SCC}}
\newcommand{\drssccb}{\kwnospace{1PB}\textrm{-}\kw{SCC}}

\newcommand{\Bdrsscc}{\kwnospace{B}\textrm{-}\kwnospace{BR'}\textrm{-}\kw{SCC}}

\newcommand{\deprtree}{depth-ranked tree\xspace}
\newcommand{\cdeprtree}{Depth-Ranked Tree\xspace}
\newcommand{\drtree}{\kwnospace{BR}\textrm{-}\kw{Tree}}
\newcommand{\drplustree}{\kwnospace{BR}$^+$\textrm{-}\kw{Tree}}
\newcommand{\drscc}{\kwnospace{2P}\textrm{-}\kw{SCC}}
\newcommand{\updatedrank}{\kwnospace{update}\textrm{-}\kw{drank}}

\newcommand{\drtreeconstruct}{\kwnospace{Tree}\textrm{-}\kw{Construction}}
\newcommand{\drtreesearch}{\kwnospace{Tree}\textrm{-}\kw{Search}}
\newcommand{\depthrerank}{\kw{pushdown}}
\newcommand{\itrerank}{\kwnospace{iterative}\textrm{-}\kw{rerank}}
\newcommand{\drr}{\Downarrow}
\newcommand{\reach}{\kw{Rset}}
\newcommand{\earlyrejection}{\kwnospace{early}\textrm{-}\kw{rejection}}
\newcommand{\earlyacceptance}{\kwnospace{early}\textrm{-}\kw{acceptance}}
\newcommand{\greduce}{\earlyacceptance}
\newcommand{\drea}{\kwnospace{1P}\textrm{/}\kw{ER}}
\newcommand{\myinf}{\kw{INF}}

\newcommand{\gcloud}{\kw{GCloud}\xspace}
\newcommand{\degree}{\kw{Degree}\xspace}
\newcommand{\subgraph}{\kw{Subgraph}\xspace}
\newcommand{\pagerank}{\kw{PageRank}\xspace}
\newcommand{\bfs}{\kw{BFS}\xspace}
\newcommand{\keysearch}{\kw{KWS}\xspace}
\newcommand{\cc}{\kw{CC}\xspace}
\newcommand{\msf}{\kw{MSF}\xspace}
\newcommand{\dmax}{\kw{rmax}}
\newcommand{\mystar}{\kw{star}\xspace}
\newcommand{\twitter}{{\sl{Twitter-2010}}\xspace}
\newcommand{\friendster}{{\sl{Friendster}}\xspace}
\definecolor{lgray}{gray}{0.85}
\definecolor{llgray}{gray}{0.9}
\newcommand{\mycc}{CC\xspace}
\newcommand{\myccs}{CCs\xspace}
\newcommand{\mymsf}{MSF\xspace}
\newcommand{\oneroundmsf}{\kw{OneRoundMSF}}
\newcommand{\multiroundmsf}{\kw{MultiRoundMSF}}

\newcommand{\hashtomin}{\kw{HashToMin}\xspace}
\newcommand{\hashgreatertomin}{\kw{HashGToMin}\xspace}
\newcommand{\pramsimulation}{\kwnospace{PRAM}\textrm{-}\kw{Simulation}\xspace}

\newcommand{\pagerankpig}{\kwnospace{PageRank}\textrm{-}\kw{Pig}\xspace}
\newcommand{\bfspig}{\kwnospace{BFS}\textrm{-}\kw{Pig}\xspace}
\newcommand{\keysearchpig}{\kwnospace{KWS}\textrm{-}\kw{Pig}\xspace}

\newcommand{\ttwig}{\kwnospace{TwinTwig}\xspace}
\newcommand{\ttwigs}{\kwnospace{TwinTwig}s\xspace}
\newcommand{\ttjoin}{\kwnospace{TwinTwig}\kw{Join}}
\newcommand{\sdec}{\kwnospace{SDEC}\xspace}
\newcommand{\subgenum}{\kwnospace{SubgraphEnum}\xspace}
\newcommand{\mymap}{\kwnospace{map}\xspace}
\newcommand{\myreduce}{\kwnospace{reduce}\xspace}
\newcommand{\cascadejoin}{\kwnospace{Edge}\kw{Join}}
\newcommand{\starjoin}{\kwnospace{Star}\kw{Join}}
\newcommand{\multiwayjoin}{\kwnospace{Multiway}\kw{Join}}
\newcommand{\cost}{\kwnospace{cost}\xspace}
\newcommand{\mysize}{\kwnospace{card}\xspace}
\newcommand{\er}{\kwnospace{ER}\xspace}
\newcommand{\optdec}{\kwnospace{Optimal}\textrm{-}\kwnospace{Decomp}\xspace}

\newcommand{\ttone}{\kwnospace{TT1}\xspace}
\newcommand{\tttwo}{\kwnospace{TT2}\xspace}
\newcommand{\ttthree}{\kwnospace{TT3}\xspace}

\newcommand{\alEdge}{\kwnospace{Edge}\xspace}
\newcommand{\alMul}{\kwnospace{Mul}\xspace}
\newcommand{\alStar}{\kwnospace{Star}\xspace}

\newcommand{\alTTBO}{\kwnospace{TTBS}\xspace}
\newcommand{\alTTNLB}{\kwnospace{TTOA}\xspace}
\newcommand{\alTTLB}{\kwnospace{TTLB}\xspace}
\newcommand{\alTTFil}{\kwnospace{TT}\xspace}

\newcommand{\reffig}[1]{Fig.~\ref{fig:#1}}
\newcommand{\refsec}[1]{Section~\ref{sec:#1}}
\newcommand{\reftable}[1]{Table~\ref{tab:#1}}
\newcommand{\refalg}[1]{Algorithm~\ref{alg:#1}}
\newcommand{\refdef}[1]{Definition~\ref{def:#1}}
\newcommand{\refthm}[1]{Theorem~\ref{thm:#1}}
\newcommand{\reflem}[1]{Lemma~\ref{lem:#1}}
\newcommand{\refex}[1]{Example~\ref{ex:#1}}

\makeatletter
\newcommand{\rmnum}[1]{\romannumeral #1}
\newcommand{\Rmnum}[1]{\expandafter\@slowromancap\romannumeral #1@}
\makeatother

\newcommand{\enumall}{\kwnospace{Enum}\kw{All}}
\newcommand{\enumsub}{\kwnospace{Enum}\kw{Sub}}
\newcommand{\cliqueall}{\kwnospace{Clique}\kw{All}}
\newcommand{\cliquesub}{\kwnospace{Clique}\kw{Sub}}
\newcommand{\maxcover}{\kwnospace{Max}\kw{Cover}}
\newcommand{\enumk}{\kwnospace{Enum}\kw{K}}
\newcommand{\priv}{\kw{priv}}
\newcommand{\cov}{\kw{cov}}
\newcommand{\enumkbasic}{\kwnospace{Enum}\kwnospace{K}\kw{Basic}}
\newcommand{\enumkopt}{\kwnospace{Enum}\kwnospace{K}\kw{Opt}}
\newcommand{\candmaintainbasic}{\kwnospace{Cand}\kwnospace{Maintain}\kw{Basic}}
\newcommand{\candmaintain}{\kwnospace{Cand}\kw{Maintain}}
\newcommand{\pnpindex}{\kwnospace{PNP}\textrm{-}\kw{Index}}
\newcommand{\rcov}{\kw{rcov}}
\newcommand{\rpriv}{\kw{rpriv}}
\newcommand{\insertc}{\kw{Insert}}
\newcommand{\deletec}{\kw{Delete}}
\newcommand{\calp}{\kwnospace{Cal}\kw{P}}

\newcommand{\myscore}{\kw{score}}
\newcommand{\initk}{\kw{InitK}}
\newcommand{\globalpruning}{\kw{GlobalPruning}}
\newcommand{\localpruning}{\kw{LocalPruning}}
\newcommand{\cliquek}{\kw{CliqueK}}
\newcommand{\mycolor}{\kw{color}}
\newcommand{\mycore}{\kw{core}}
\newcommand{\cliquegreedy}{\kw{CliqueGreedy}}

\newcommand{\sops}{\kw{SOPS}}
\newcommand{\gops}{\kw{GOPS}}
\newcommand{\sieve}{\kw{SIEVE}}
\newcommand{\enumkglobal}{\kw{Global}}
\newcommand{\enumklocal}{\kw{Local}}

\newcommand{\dsgoogle}{\textit{Google}\xspace}
\newcommand{\dsskitter}{\textit{Skitter}\xspace}
\newcommand{\dsyoutube}{\textit{Youtube}\xspace}
\newcommand{\dspokec}{\textit{Pokec}\xspace}
\newcommand{\dswiki}{\textit{Wiki}\xspace}
\newcommand{\dsundirecteda}{Different $\epsilon$ ($\mu$ = 2)\xspace}
\newcommand{\dsundirectedb}{Different $\epsilon$ ($\mu$ = 3)\xspace}
\newcommand{\dsundirectedc}{Different $\epsilon$ ($\mu$ = 4)\xspace}
\newcommand{\dsundirectedd}{Different $\mu$ ($\epsilon$ = 0.6)\xspace}

\newcommand{\dsdirecteda}{Different $\epsilon_f$ ($\mu$=2, $\epsilon_c$=0.45)\xspace}
\newcommand{\dsdirectedb}{Different $\epsilon_c$ ($\mu$=2, $\epsilon_f$=0.45)\xspace}
\newcommand{\dsdirectedc}{Different $\epsilon_f$ ($\mu$=3, $\epsilon_c$=0.45)\xspace}
\newcommand{\dsdirectedd}{Different $\epsilon_c$ ($\mu$=3, $\epsilon_f$=0.45)\xspace}
\newcommand{\dsdirectede}{Different $\epsilon_c$ and $\epsilon_f$($\mu$ = 2)\xspace}
\newcommand{\dsdirectedf}{Different $\mu$ ($\epsilon_c$=$\epsilon_f$=0.45)\xspace}

\newcommand{\dswebgoogle}{web-Google\xspace}
\newcommand{\dsLiveJournal}{LiveJournal1\xspace}
\newcommand{\dswebbase}{webbase-2001\xspace}
\newcommand{\dsuk}{uk-2002\xspace}

\newcommand{\dscaseaa}{$\mu$ = 4, $\epsilon_c$ = 0.22, $\epsilon_f$ = 0.18\xspace}
\newcommand{\dscaseab}{$\mu$ = 4, $\epsilon$ = 0.52\xspace}

\newcommand{\dscaseba}{$\mu$ = 3, $\epsilon_c$ = 0.3, $\epsilon_f$ = 0.3\xspace}
\newcommand{\dscasebb}{$\mu$ = 3, $\epsilon$ = 0.4\xspace}

\newcommand{\dscaseca}{$\mu$ = 4, $\epsilon_c$ = 0.24, $\epsilon_f$ = 0.24\xspace}
\newcommand{\dscasecb}{$\mu$ = 4, $\epsilon$ = 0.41\xspace}

\newcommand{\dsuntri}{Undirected\xspace}
\newcommand{\dscycletri}{Cycle\xspace}
\newcommand{\dsflowtri}{Flow\xspace}

\newcommand{\nbr}{\kw{nbr}\xspace}
\newcommand{\nbrin}{\kw{nbr^-}}
\newcommand{\nbrout}{\kw{nbr^+}}
\newcommand{\mydeg}{\kw{deg}}
\newcommand{\mydegin}{\kw{deg^-}}
\newcommand{\mydegout}{\kw{deg^+}}
\newcommand{\scan}{\kw{SCAN}}
\newcommand{\dscan}{\kw{DSCAN}}
\newcommand{\clusteringq}{$Q_{\epsilon_c, \epsilon_f, \mu}$\xspace}
\newcommand{\dscanindex}{\kwnospace{DSCAN}\textrm{-}\kw{Index}}
\newcommand{\dscanquery}{\kwnospace{DSCAN}\textrm{-}\kw{Query}}
\newcommand{\dscanquerynospace}{\kwnospace{DSCAN}\textrm{-}\kwnospace{Query}}

\newcommand{\hcst}{\kwnospace{HC}\textrm{-}\kwnospace{s}\textrm{-}\kw{t}}
\newcommand{\hcstp}{\kwnospace{HC}\textrm{-}\kwnospace{s}\textrm{-}\kw{t~path}}
\newcommand{\hcsp}{\kwnospace{HC}\textrm{-}\kwnospace{s}\kw{~path}}
\newcommand{\hcstps}{\kwnospace{HC}\textrm{-}\kwnospace{s}\textrm{-}\kw{t~paths}}
\newcommand{\bcdfs}{\kwnospace{BC}\textrm{-}\kw{DFS}}
\newcommand{\disbfs}{\kw{DisBFS}}
\newcommand{\bfsenum}{\kw{BiGJoin}}
\newcommand{\hpindex}{\kwnospace{HP}\textrm{-}\kw{Index}}
\newcommand{\dishpindex}{\kwnospace{DisHP}\textrm{-}\kw{Index}}
\newcommand{\hpi}{\kw{HPI}}
\newcommand{\dishpi}{\kw{DisHPI}}
\newcommand{\centralized}{\kw{DFSEnum}}
\newcommand{\hybridsearch}{\kw{HybridEnum}}
\newcommand{\hybridenum}{\kw{HybridEnum^+}}
\newcommand{\pextend}{\kw{PrunExt}}
\newcommand{\backprop}{\kw{BackProp}}
\newcommand{\jump}{\kw{Jump}}
\newcommand{\level}{\kw{level}}
\newcommand{\degout}{\kw{deg^+}}
\newcommand{\degin}{\kw{deg^-}}
\newcommand{\flowctrl}{\kw{FlowCtrl}}
\newcommand{\concat}{\kw{ConCat}}
\newcommand{\dobfs}{\kw{DOBFS}}
\newcommand{\mobfs}{\kw{RQBFS}}
\newcommand{\msbfs}{\kw{MSBFS}}
\newcommand{\fractal}{\kw{Fractal}}
\newcommand{\tdfs}{\kw{T}-\kw{DFS}}
\newcommand{\tdfss}{\kw{T}-\kw{DFS2}}
\newcommand{\cdobfs}{\kw{DOBFS}-\kw{C}}
\newcommand{\cmobfs}{\kw{RQBFS}-\kw{C}}
\newcommand{\cmsbfs}{\kw{MSBFS}-\kw{C}}
\newcommand{\mhcstp}{\kwnospace{Multi}\textrm{-}\kwnospace{HC}\textrm{-}\kwnospace{s}\textrm{-}\kw{t~path}}
\newcommand{\shortcutg}{\kw{Shortcut}-\kw{Graph}}
\newcommand{\mic}{\kw{ConstructIndex}}
\newcommand{\sharedetect}{\kw{HCShareDetect}}
\newcommand{\querycluster}{\kw{Cluster}\kw{Query}}
\newcommand{\overlapextract}{\kw{Detect}\kw{CommonQuery}}
\newcommand{\plangen}{\kw{Plan}\kw{Gen}}
\newcommand{\mpathenum}{\kw{Multi}\kw{PathEnum}}
\newcommand{\pathenum}{\kw{BasicEnum}}
\newcommand{\batchenum}{\kw{BatchEnum}}
\newcommand{\pathenump}{\kw{BasicEnum^+}}
\newcommand{\batchenump}{\kw{BatchEnum^+}}

\newcommand{\mqo}{\kw{MQO}}
\newcommand{\order}{\kw{GeneratePlan}}
\newcommand{\batchpe}{\kw{BatchHcPE}}
\newcommand{\peindex}{\kw{PE}-\kw{Index}}
\newcommand{\peindices}{\kw{PE}-\kw{Indices}}
\newcommand{\beindex}{\kw{BE}-\kw{Index}}
\newcommand{\senum}{\kw{BasicEnum}}
\newcommand{\sharedenum}{\kw{SharedEnum}}
\newcommand{\seqs}{\kw{SeqDFS}}
\newcommand{\seqb}{\kw{SeqJOIN}}
\newcommand{\seqo}{\kw{SeqOpt}}
\newcommand{\penum}{\kw{PRAMEnum}}
\newcommand{\paraenum}{\kw{BasePEnum}}
\newcommand{\heft}{\kw{HEFT}}
\newcommand{\scenum}{\kw{SchedulePEnum}}
\newcommand{\ecost}{\kwnospace{ext}\textrm{-}\kw{cost}}
\newcommand{\ccost}{\kwnospace{concat}\textrm{-}\kw{cost}}

\newcommand{\commonenum}{\kw{common} \kwnospace{HC}\textrm{-}\kwnospace{s}\textrm{-}\kw{t~paths}}
\newcommand{\subquery}{\kw{maximal~common} \kwnospace{HC}\textrm{-}\kwnospace{s}\textrm{-}\kw{t~path~subquery}}

\newcommand{\mqopath}{$\kwnospace{MQO}_{\kwnospace{HC}\textrm{-}\kwnospace{s}\textrm{-}\kwnospace{t}\textrm{-}\kw{path}}$}

\title{Batch  Hop-Constrained s-t Simple Path Query Processing  in Large   Graphs}


%

\author{
Long Yuan$^{\natural}$, Kongzhang Hao$^{\natural}$, Xuemin Lin$^{\ddag}$, Wenjie Zhang$^{\S}$ \\
\fontsize{9}{9}\selectfont\itshape
$^{\natural}$Nanjing University of Science and Technology, China,  $^{\ddag}$Shanghai Jiao Tong University, China
$^{\S}$The University of New South Wales, Australia\\
\fontsize{9}{9}\selectfont\ttfamily\upshape
longyuan@njust.edu.cn, 
haokongzhang@gmail.com,
xuemin.lin@sjtu.edu.cn,
zhangw@cse.unsw.edu.au
}

\newcommand\blfootnote[1]{%
  \begingroup
  \renewcommand\thefootnote{}\footnote{#1}%
  \addtocounter{footnote}{-1}%
  \endgroup
}

\maketitle

\begin{abstract}
Hop-constrained $s$-$t$ simple path ($\hcstp$) enumeration is a fundamental problem in graph analysis. Existing solutions for this problem focus on optimizing the processing performance of a single query. However, in practice, it is  more often that multiple $\hcstp$  queries are issued simultaneously and  processed as a batch. Therefore, we study the problem of batch  \hcstp query processing in this paper and aim to compute the results of all  queries concurrently and efficiently as a batch. To achieve this goal, we first propose the concept of \hcsp query which can precisely characterize the common computation among different queries.  We then devise a  two-phase \hcsp query detection algorithm to identify the common \hcsp queries for the given \hcstp queries. Based on the detected \hcsp queries, we further devise an efficient \hcstp enumeration algorithm in which the common computation represented by \hcsp queries are effectively shared.  We conduct extensive experiments on real-world graphs and the experimental results demonstrate that our proposed algorithm is efficient and scalable regarding  processing multiple \hcstp  queries in large graphs at billion-scale.
\blfootnote{$^*$Long Yuan and Kongzhang Hao are the joint first authors. }
\end{abstract}




\section{Introduction}
\label{sec:intro}

Graphs have been widely used to represent the relationships of entities in many areas including social networks, web graphs, and biological networks. With the proliferation of graph applications, research efforts have been devoted to many fundamental problems in analyzing graphs \cite{lu01, lu02, chengfei16,xiaofei19, ye20, chen2020efficient,wook16,chen2021higher, chen2022balanced,wang2023towards}. Among them, the problem of hop-constrained s-t simple path enumeration has received considerable attention \cite{Ri14, Gr18, Qi18, Pe19, pathenum,hao2021distributed}.  Given an unweighted directed graph $G$, a source vertex $s$, a target vertex $t$, and a hop constraint $k$, hop-constrained s-t simple path (\hcstp for short) enumeration computes all the simple paths (i.e., paths without repeated vertices) from $s$ to $t$ such that the number of hops in each path is not larger than $k$.

\stitle{Applications.} \hcstp enumeration can be used in many applications, for example:

 \noindent$\bullet$  \emph{Fraud detection in E-commerce transaction networks}. A cycle in an E-commerce  transaction network is a strong indication of a fraudulent activity \cite{Yu07}. According to a recent paper by the Alibaba group \cite{Qi18},   \hcstp enumeration is used to report all newly formed cycles to detect fraudulent activities when a new transaction is submitted from account $t$ to account $s$ in its E-commerce network.
 

 \noindent$\bullet$  \emph{Pathway queries in biological networks.}  Pathway queries are a fundamental tool in biological networks analytics \cite{Kr03,Le05}.  \cite{Le05} shows that \hcstp enumeration is one of the most important pathway queries in figuring out the chains of interactions between multiple pairs of substances.

 \noindent$\bullet$  \emph{Knowledge graph completion}.  Knowledge graphs (KGs) are widely used in a large number of applications. Because KGs are generally incomplete, it is commonly necessary to predict the missing relations, which is known as the problem of knowledge graph completion. Knowledge graph completion methods generally enumerate the paths between two entities to train models to predict the relationship, where the entities connected by many short paths have a higher tendency to be related, e.g., \cite{sh16, sh17}, suggesting the utility of hop constraints in this setting.
 


 
 \stitle{Motivation.} Although considerable algorithms for \kwnospace{HC}\textrm{-}\kwnospace{s}\textrm{-}\kw{t} path enumeration have been proposed in the literature \cite{Ri14, Gr18, Qi18, Pe19, pathenum}, all of these algorithms focus on "a-query-at-a-time" scenario and assume queries are isolated and evaluated independently. However, in real application scenarios of \hcstp queries, it is also  often that multiple queries are issued simultaneously and processed as a batch. For example, in the fraud detection application, a large number of transactions could occur within a short time period and these transactions are submitted to the system in a batch. Regarding knowledge graph prediction, missing relations exist between a large number of entity pairs, which need to be predicted together. As a result, multiple \kwnospace{HC}\textrm{-}\kwnospace{s}\textrm{-}\kw{t} path queries need to be processed at the same time.   
 
 Motivated by this, we study the problem of  batch \hcstp  query processing  in this paper. Specifically, given an unweighted directed graph $G$ and a set of \hcstp enumeration queries $Q = \{q_0, \dots, q_n\}$, we aim  to efficiently enumerate the  \hcstps for all the queries in $Q$. 
 
 
\stitle{Challenges.} {To address this problem, we can treat  queries in $Q$ individually, and process each single query using the state-of-the-art \hcst path enumeration algorithm sequentially or deploy more servers to process these queries in parallel. However, the success of batch processing optimization (also known as multiple query optimization) for other problems such as \emph{SQL} \cite{Se88, Se90, Fi82,DBLP:conf/bigdataconf/TuEXC22}, \emph{SPARQL} \cite{Le12, lumulti,DBLP:journals/vldb/AliSYHN22},  subgraph isomorphism \cite{Re16},  regular path queries \cite{DBLP:conf/icde/Abul-Basher17} and shortest path queries \cite{li2020fast}  shows the possibility of improving the performance of sequential batch query processing by sharing common computation across queries, which implies that processing these \hcstp enumeration  queries separately or directly in parallel may miss potential performance improvement by sharing the common computation, and thus leads to resource inefficiency. Moreover, in real application scenarios of \hcstp queries, it is very likely that the queries in a batch share common computation. For example, in knowledge graph completion,  two queries related to the same vertex could involve similar paths.   This inspires us to optimize the processing of multiple \hcstp queries.  Specifically, for a given set of \hcstp queries, (1) when there is a significant amount of common computation among the queries, we want to make full use of the common computation to accelerate the whole batch process; (2) when there is little or no common computation among the queries, we want the overhead caused by computation sharing to be as low as possible, namely the total processing time should be similar to sequential processing.}

Unfortunately, finding an optimal solution that can make full use of the common computation for batch \hcstp query processing is  NP-hard as shown in \refsec{pre}, which means it is difficult to design an efficient algorithm that can achieve our goals discussed above. A natural question that arises is: can we extend the existing batch processing techniques to our problem? No matter how desirable, the answer is negative. Taking the batch subgraph isomorphism query processing as an example, which is the closest problem to ours in the literature (other related problems are discussed in \refsec{related}).  Given a data graph and a set of query graphs, batch subgraph isomorphism query processing aims to find all subgraphs of the data graph that are isomorphic to one of the query graphs. Ren et al. \cite{Re16} proposes an effective computation sharing strategy by detecting and extracting the overlaps among the query graphs and reusing the partial results which are isomorphic to the overlaps to speed up the processing of all these queries as a whole. Obviously,  the overlaps among the queries play the key role in the success of the proposed method. However, for our batch  \hcst path  query processing, we cannot even find such a counterpart of the overlaps among the given queries since a \hcstp enumeration query only contains three parameters: the source vertex $s$, the target vertex $t$, and a hop constraint $k$, which are uninformative compared with the massive number of  returned   \hcst paths. Therefore, we have to make a fresh start and develop an efficient and scalable algorithm tailored for the batch \hcst path query processing in large graphs. The algorithm should effectively share the computation among the queries, and consequently reduce the overall processing time. 


Nevertheless, it is challenging to develop  an algorithm which can achieve the above goals at the same time for the following reasons: (1) how to design such  a  structure that can precisely characterize the common computation among different queries? (2) based on the designed structure, how to efficiently detect the corresponding structure such that the performance improvement obtained due to computation sharing outweighs the overhead caused by detecting the structure? (3) with the detected structure, how to enumerate the \hcstps such that the common computation can be maximally shared, and thus improve the whole processing performance?  

\vspace{-1mm}

\stitle{Contributions.} The contributions  are summarized as follows: 

\vspace{-1mm}

\sstitle{(A) The first work to study the batch   \hcstp query processing.} In this paper, we aim to efficiently enumerate  the \hcstps for a given set of  \hcst path enumeration queries. To the best of our knowledge, this is the first work to study the batch \hcstp  query processing in large graphs.

\vspace{-1mm}

\sstitle{(B) Efficient and scalable algorithm for multiple \hcst path queries.} 
By revisiting the state-of-the-art \hcstp enumeration algorithm for a single query, we first propose the concept of \hcsp query based on which the common computation among different queries in $Q$ can be shared. After this, we propose a two-phase algorithm to detect the common \hcsp queries among the queries. Following the detected \hcsp queries, we devise an efficient batch \hcstp processing algorithm which can significantly improve the enumeration performance by reusing the common computation.


\vspace{-1mm}

\sstitle{(C) Extensive performance studies on real-world datasets.} We conduct extensive performance studies on  real-world graphs with various graph properties. The experimental results demonstrate that our proposed algorithm is efficient and scalable regarding processing multiple \hcstp enumeration queries in large graphs with billion-scale edges. 



\section{Preliminaries}
\label{sec:pre}
Let $G = (V, E)$ denote an unweighted directed graph, where $V(G)$ is the set of vertices and $E(G)$ is a set of directed edges. We use $n$ and $m$ to denote the number of vertices and edges, respectively. For a vertex $v \in V(G)$, we use $G.\nbrin(v)$/$G.\nbrout(v)$ to denote the in-neighbors/out-neighbors of $v$ in $G$. Given a graph $G$, the reverse graph of $G$, denoted by $G_r = (V, E_r)$, is the graph generated by reversing the direction of all edges in $G$. A path from  vertex $u$ to  vertex $v$, denoted by $p(u, v)$, is a sequence of vertices $\{u = v_0, v_1, ..., v_h = v\}$ such that $(v_{i-1}, v_i) \in E(G)$ for every $1 \leq i < h$.  Given two paths $p_A$ and $p_B$, $p_A$ is a partial path of $p_B$ if $p_A$ makes up part of $p_B$, denoted by $p_A \subseteq p_B$. A simple path is a loop-free path where there are no repetitions of vertices and edges. By $|p|$ and $p[i]$, we denote the length of path $p$ and the $i^{th}$ vertex of $p$, respectively. { Given two vertices $u$ and $v$, the shortest distance from $u$ to $v$, denoted by $\kw{dist}_G(u, v)$, is the length (i.e., the number of hops in this paper) of the shortest path from $u$ to $v$ on $G$, we omit $G$ in the notations  when the context is self-evident. Given a pre-defined hop constraint $k$, we say that a path $p$ is a hop-constrained path if $|p| \le k$.  We call a traversal on $G$ as a forward search and on $G_r$ as a backward search.  Given an unweighted directed graph $G$, a source vertex $s$, a target vertex $t$, and a hop constraint $k$, a \hcstp enumeration query $q(s, t, k)$ aims to enumerate all simple paths from $s$ to $t$ with the number of hops not larger than $k$. The results of a \hcstp query $q(s, t, k)$ are denoted as $P(q)$ or $P(q(s, t, k))$. Letting $p$ be a path in $P(q(s, t, k))$,  we call the sub-path of $p$ starting from $s$ the prefix of $q$.}

\begin{figure}[htb]
  \centering
  \includegraphics[width=0.9\columnwidth]{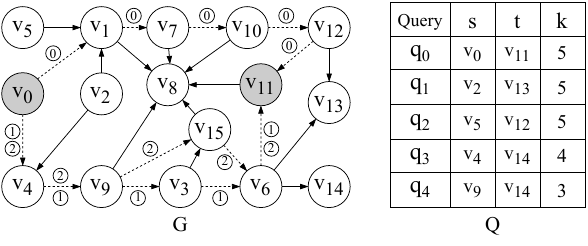}
        \vspace{-3mm}
  \caption{A graph $G$ and multiple \hcstps queries $Q$}
  \label{fig:graph}
      \vspace{-5mm}
\end{figure}

\stitle{Problem statement.} Given an unweighted directed graph $G$ and a batch of \hcstp enumeration queries $Q = \{q_0, ..., q_n\}$, we aim to \textit{efficiently} enumerate the \hcstps of all the queries in $Q$ on graph $G$. 

\begin{example}
Consider an unweighted directed graph $G$ and a set of \hcstp queries $Q$ shown  in \reffig{graph}. For $q_0$, three \hcstps can be found, namely $p_0 = (v_0, v_1, v_7, v_{10}, v_{12}, v_{11})$, $p_1 = (v_0, v_4, v_9, v_3, v_6, v_{11})$ and $p_2 = (v_0, v_4, v_9, v_{15}, v_6, v_{11})$, which are demonstrated by the dashed arrows and the id labels in \reffig{graph}. The \hcstps for the remaining queries in $Q$ can be obtained similarly.
\end{example}

Finding an optimal solution for batch \hcstp query processing is NP-hard, which can be easily proved based on the NP-hardness of the multiple query optimization problem for relational queries. Given a set of relational queries,  the multiple query optimization problem for relational queries aims to find a global access plan such that the cost to answer these queries is minimum by sharing common data, and T. Sellis et al. \cite{sellis1990multiple} proves the NP-hardness of this problem. Meanwhile, the relational queries and the \hcstp queries can be equivalently transformed  as shown in  \cite{pathenum}. Following  \cite{sellis1990multiple}, we can derive the NP-hardness of batch \hcstp query processing. Due to the NP-hardness of the problem, we aim to design a sub-optimal but practically effective and efficient algorithm that can accelerate the batch \hcstp query processing.

\begin{figure}[htb]
  \centering
  \includegraphics[width=0.8\columnwidth]{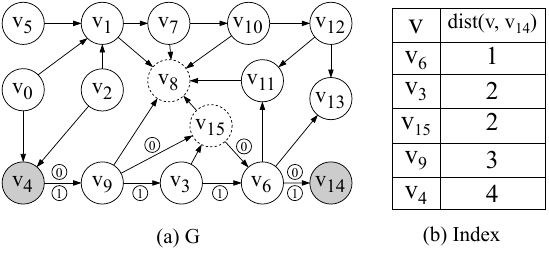}
      \vspace{-3mm}

  \caption{Procedures of \pathenum}
  \label{fig:baseline}
    \vspace{-5mm}
\end{figure}

\begin{figure*}[!htb]
 \begin{subfigure}[b]{0.6\columnwidth}
  \centering
  \includegraphics[width=\columnwidth]{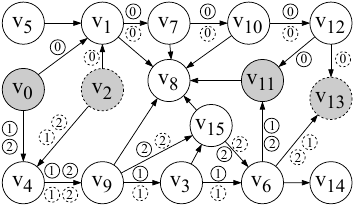}
      \vspace{-5mm}
  \caption{G}
  \end{subfigure}
  \hspace{0.2cm}
   \begin{subfigure}[b]{0.47\columnwidth}
  \centering
  \includegraphics[width=\columnwidth]{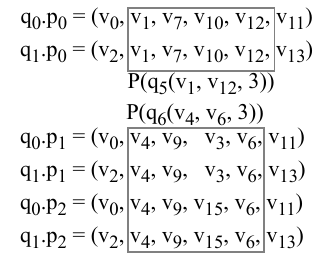}
      \vspace{-5mm}
      \hspace{4mm}
  \caption{\hcstps}
  \end{subfigure}
  \hspace{0.2cm}
    \begin{subfigure}[b]{0.9\columnwidth}
         \centering
         \includegraphics[width=\columnwidth]{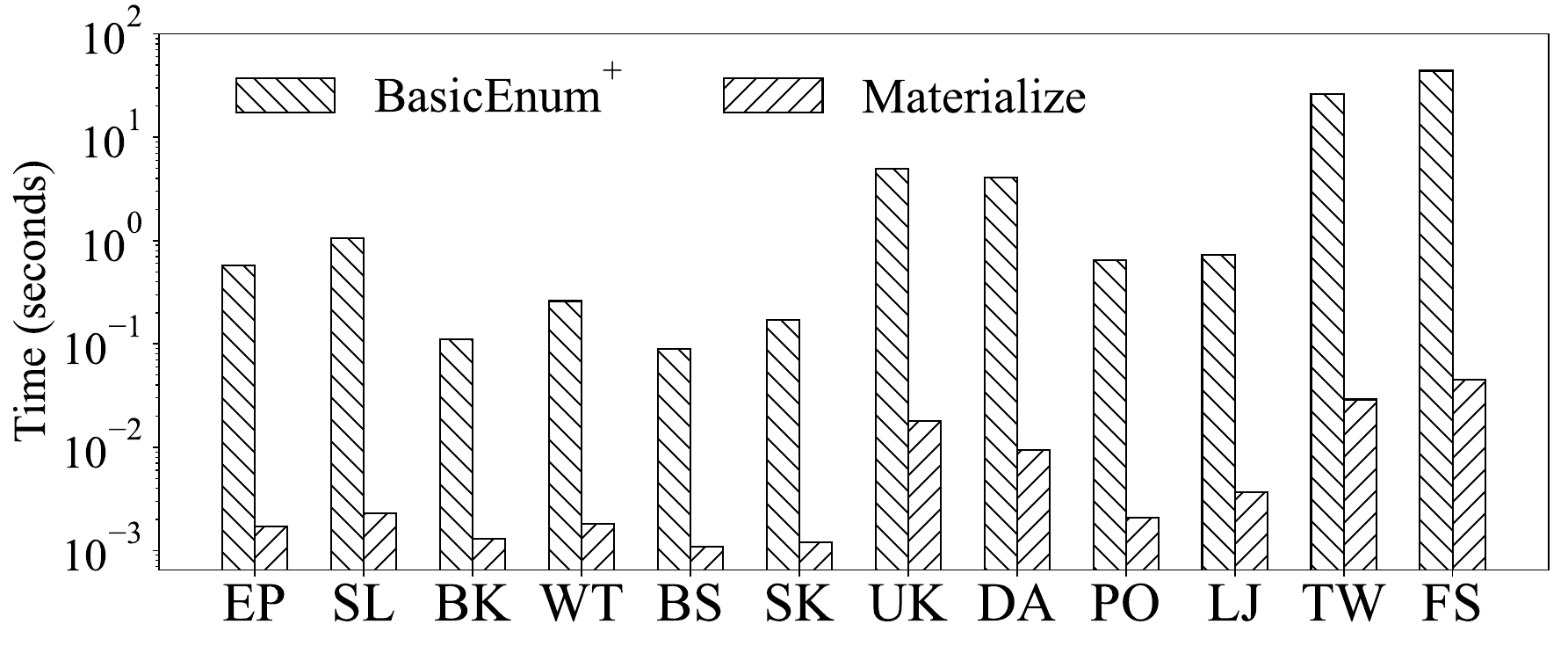}
         \vspace{-5mm}
         \caption{Materialization time}
         \label{fig:commonenummat}
     \end{subfigure}
     \vspace{-5mm}
     \caption{Main observation of our approach}
  \label{fig:commonenum}
      \vspace{-6mm}
\end{figure*}

\section{Baseline Solution}
\label{sec:exist}

In this section, we introduce the state-of-the-art algorithm, \kw{PathEnum} \cite{pathenum}, for a single \hcstp enumeration query and present a straightforward  baseline solution for batch \hcstp enumeration processing  based on \kw{PathEnum}, which paves the way for our optimized batch query processing. The main idea of \kw{PathEnum} is based on the following lemma:



\begin{lemma}
\label{lem:pe}
\cite{pathenum} Given a vertex $v \in V(G)$, if there exists a path $p$ from $s$ to $t$ with $|p| \le k$, then for any $p[i] = v$ where $0 \le i \le |p|$, $\kw{dist}(s,v) \le i$ and $\kw{dist}(v,t) \le k-i$.
\end{lemma}

According to \reflem{pe}, given a partial path $p$ where $v$ is the ending vertex, \kw{PathEnum} only considers the neighbors $v'$ of $v$ such that $|p| + \kw{dist}_G(v', t) < k$, or $|p| + \kw{dist}_{G_r}(v', s) < k$ if it searches in the backward direction based on $G_r$ during the \hcstp enumeration, as other neighbors cannot lead to paths satisfying the hop constraint $k$. Based on this observation, \kw{PathEnum} treats $\kw{dist}_G(s,v)$ and $\kw{dist}_{G_r}(t, v)$ with value not greater than $k$ for $v \in V(G)$ as the key index structure (The real index structure of \kw{PathEnum} is slightly different, but the key idea is the same)  and uses the index to prune unnecessary exploration during the  enumeration. Moreover, as bidirectional search could improve the enumeration performance, \kw{PathEnum} conducts a forward search from $s$ on $G$ and a backward search from  $t$ on $G_r$ concurrently, and concatenates the paths explored during the bidirectional search to obtain the final results by  the path concatenation $\oplus$ operator, which is defined as follows:   

\begin{definition}
\label{def:join}
\textbf{(Path Concatenation $\oplus$)} Given two sets of paths $\mathcal{P}_A$ and $\mathcal{P}_B$, $\mathcal{P}_A \oplus \mathcal{P}_B$ generates the set of paths $p$ by hash joining $p_A \in \mathcal{P}_A$ and $p_b \in \mathcal{P}_B$ on $p_A[|p_A|] = p_B[0]$.
\end{definition}

\begin{algorithm}[!t]
  \SetAlgoLined\DontPrintSemicolon
  \SetKwFunction{proc}{proc}
 \nl $S \leftarrow \bigcup_{q \in Q} \{q.s\}$; $T \leftarrow  \bigcup_{q \in Q} \{q.t\}$
  
  \nl construct index by  multi-source \kwnospace{BFS}s from $S$ and $T$;\\
   \nl \ForEach{$q \in Q$}{
  \nl $P_f \leftarrow \emptyset$; $P_b \leftarrow \emptyset$;\\
  \nl \kwnospace{Search}$(G, P_f, q.s, q.t, \ceil{q.k/2})$;\\
  \nl \kwnospace{Search}$(G_r, P_b, q.t, q.s, \floor{q.k/2})$;\\
  \nl \ForEach{$p \in P_f \oplus P_b$}{
  \nl \textbf{if} $p$ has no duplicated vertex \textbf{then} \kwnospace{Output} $p$;
  }
  }

  \nl \SetKwProg{myproc}{Procedure}{}{}
  \myproc{\kwnospace{Search}$(G, P, p, v, k)$}{
  \nl $v' \leftarrow p[|p|]$; $P.\kw{add}(p)$;\\
  \nl \textbf{if} $|p| = k$ \kw{or} $k = 0$ \textbf{then}  \KwRet ; \\
  \nl \ForEach{$v'' \in \nbrout(v')$ s.t. $|p| + \kw{dist}_G(v'', v) < k$}{
  \nl \textbf{if} $v'' \not\in p$ \textbf{then} \kwnospace{Search}($G, P, p \bigcup \{v''\}, v, k$);
  }}
  \caption{\kwnospace{BasicEnum}($G, Q$)}
  \label{alg:base}
\end{algorithm}

\stitle{Algorithm.} Based on \kw{PathEnum}, the baseline solution for batch \hcstp query processing is as follows: since the index for a single query $q(s, t, k)$ can be constructed by two \kwnospace{BFS}s from $s$ and $t$ on $G$ and $G_r$, the index for the batch queries $Q$ can be constructed by the state-of-the-art multi-source \kwnospace{BFS}s from $S$ and $T$, where $S$ represents the source vertex set $\cup_{q \in Q} \{q.s\}$ and  $T$ represents the target vertex set $\cup_{q \in Q} \{q.t\}$. Based on the index,  $Q$ can be considered as a series of individual queries, and each query can be processed separately using \kw{PathEnum}. It is immediate that all \hcstps of the queries in $Q$ can be correctly enumerated by this approach.  \refalg{base} illustrates the baseline solution \pathenum for batch \hcstp query processing. 

For the batch of queries  $Q$, \pathenum first performs two multi-source \kwnospace{BFS}s from $S$ and $T$, respectively, to compute $\kw{dist}_G(s, v)$ and $\kw{dist}_{G_r}(t, v)$ for all $v \in V(G)$, where $s \in \cup_{q \in Q} \{q.s\}$, $t \in \cup_{q \in Q} \{q.t\}$ (lines 1-2) (for a query $q(s, t, k)$, only the entities with $\kw{dist}_G(s, v) \leq k$ and $\kw{dist}_{G_r}(t, v) \leq k$ are stored in the index, the distance for the remaining entities are treated as $\infty$).  After this, for each query $q \in Q$, two empty sets $P_f$ and $P_b$ are created to store the \hcstps found during the search in the two directions (lines 3-4). \pathenum then conducts a forward search from $q.s$ with a hop constraint of $\ceil{q.k/2}$ on $G$ and a backward search from $q.t$ with a hop constraint of $\floor{q.k/2}$ on $G_r$ (lines 5-6). The paths obtained by the forward search and backward search are concatenated by $\oplus$ (line 7), and if there is no duplicated vertex in the concatenated path, it is a \hcstp and \pathenum outputs the path (line 8).    Procedure \kw{Search} enumerates the \hcstps with a hop constraint of $k$ based on the index recursively (lines 9-13). Specifically, paths $p$ with length smaller than $k$ are added into $P$ (lines 10-11).  For the out-neighbor $v''$ of $v'$  which meets the hop constraint and has not been explored in $p$, \pathenum adds $v''$ in  $p$ and continues the search (line 12-13). {Note that for the special hop constraint $1$,  \refalg{base}  can also return the correct results due to condition $k = 0$ checked in line 11.}




\begin{example}
Reconsider $G$ and $q_3(v_4, v_{14}, 4)$ shown in \reffig{graph}, where $q_3.s$ and $q_3.t$ are marked in grey. After running the multi-source \kwnospace{BFS}s, the entities of $\kw{dist}(v, v_{14})$ in the index are shown in \reffig{baseline} (b). Assume that the current prefix is $(v_4, v_9, v_3)$. After finishing the enumeration following $v_6$, the search backtracks and subsequently explores $v_{15}$. \pathenum  looks up the index and finds  $|(v_4, v_9, v_3)| + \kw{dist}(v_{15}, v_{14}) = 2+2 \ge q.k$. Hence, $v_{15}$ is pruned from the enumeration with the current prefix. When the search backtracks and the prefix is $(v_4, v_9)$, \pathenum similarly prunes exploration following $v_8$  as  $|(v_4, v_9)| + \kw{dist}(v_8, v_{14}) = 1 + \infty \ge q.k$. 
The pruned vertices during the enumeration are shown by the dashed circles.
\end{example}

\section{Our approach}
\label{sec:sharing}


\subsection{Overview}
\label{sec:overview}



Given a batch of \hcstp queries $Q$, \senum processes each query in $Q$ separately, which  misses the possible computation sharing opportunities among  different queries even through there exists a large amount of common computation in batch \hcstp query processing. 
Consider the \hcstp enumeration procedure regarding   $q_0(v_0, v_{11}, 5)$ and $q_1 (v_2, v_{13}, 5)$, which is shown in  \reffig{commonenum}. The $s$ and $t$ of $q_0$ are shown with the solid grey vertices and those of $q_1$ are shown with the dashed grey vertices. For query $q_0$, it has three  \hcstps, namely $P(q_0) = \{p_0  = (v_0, v_1, v_7, v_{10}, v_{12}, v_{11})$, $p_1 = (v_0, v_4, v_9, v_3, v_6, v_{11})$, $p_2 = (v_0, v_4, v_9, v_{15}, v_6, v_{11})\}$. For query $q_1$, it also has three \hcstps, namely $P(q_1) = \{p_0 = (v_2, v_1, v_7, v_{10}, v_{12}, v_{13})$, $p_1 = (v_2, v_4, v_9, v_3, v_6, v_{13})$,  $p_2 = (v_2, v_4, v_9, v_{15}, v_6, v_{13})\}$. \reffig{commonenum} (a) marks the edges that are explored by $q_0$ and $q_1$ during their corresponding \hcst path enumeration with solid and dashed path id labels, respectively,  while \reffig{commonenum} (b) lists the corresponding \hcstps for $q_0$ and $q_1$. As shown in \reffig{commonenum} (b), the \hcstp $p_0$ for $q_0$ and $p_0$ for $q_1$ share the same partial path $(v_1, v_7, v_{10}, v_{12})$. Similarly, the partial paths $(v_4, v_9, v_3, v_6)$ and $(v_4, v_9, v_{15}, v_6)$ are shared between the \hcstps $p_1$, $p_2$ for $q_0$ and $p_1$, $p_2$ for $q_1$. According to this example, it is clear that there exists a large amount of common computation between the queries in batch \hcstp query processing.




{Meanwhile, we randomly select 1000 \hcstp queries for each dataset used in our experiments, and compare the average time of processing each query by \kw{PathEnum} and the average time of directly retrieving the corresponding \hcstps of these 1000 \hcstp queries from the input graph followed by scanning them once. The results are shown in \reffig{commonenum} (c). As shown in \reffig{commonenum} (c), directly retrieving and scanning the corresponding \hcstps outperforms \kw{PathEnum} by nearly three orders of magnitude.} Therefore, if we know in advance that these \hcstps are shared between multiple \hcstps, we can enumerate and materialize them once and derive the final \hcstps based on these materialized paths. Due to the huge performance gap between computation on the fly and reusing the materialized result, the batch \hcstp query processing with sharing can  significantly improve the enumeration performance compared with \senum.



Following these observations, it is clear that the performance of batch \hcstp  query processing can be significantly improved if the common computation among different queries can be shared. To achieve this goal, we first define the common sub-structure on which our path enumeration sharing strategy relies. After this, we devise a light-weight approach that can effectively detect these common sub-structures among the queries in $Q$. Based on the detected common sub-structures, we devise an efficient batch \hcstp processing algorithm which can maximally reuse the common sub-structures, and thus improve the whole performance.

\subsection{Common Sub-Structure Detection}



As discussed in \refsec{overview},  sharing the computation of \commonenum among different \hcstp queries can significantly reduce the overall enumeration cost. However, as we do not know the queries' corresponding \hcstps in advance, it is infeasible to find such \commonenum among them. Moreover, since the problem is NP-hard as shown in \refsec{pre},  finding a solution that is able to optimally share the computation is prohibitive.  The following question arises: is it possible to design an efficient algorithm such that (1) the common sub-structures among queries can be effectively identified; (2) the identified sub-structures can be easily materialized and reused following the framework of \kw{PathEnum}? In this section, we aim to answer these questions and introduce our approach to detect the common sub-structures among queries.

Consider the enumeration procedure of \kw{PathEnum}, which treats a \hcstp as two paths from $s$/$t$ with hop constraint $\ceil{q.k/2}$/$\floor{q.k/2}$, respectively. Therefore, instead of the \hcst path, we define:

 \begin{definition}
 \textbf{(Single Source Hop-Constrained Path)}	Given a graph $G$, a source vertex $s$ and a hop constraint $k$, a single source hop-constrained path is a path starting from $s$ with length less than $k$ on $G$.
 \end{definition}

\vspace{0.5mm}

Following single source hop-constrained path, we  define:

\begin{definition}
\label{def:subquery}
\textbf{(HC-s Path Query)} Given a graph $G$, a vertex $s$ and a hop constraint $k$, a \hcsp query regarding $s$ and $k$, denoted by $\mathcal{q}_{s, k, G}$,  returns all the single source hop-constrained paths starting from $s$ with length less than $k$ on $G$.
\end{definition}

\vspace{0.5mm}

For ease of presentation, we also use $P(\mathcal{q}_{s, k, G})$ to denote the paths returned by a \hcsp query $\mathcal{q}_{s, k, G}$. Based on the procedure of the bidirectional path enumeration, it is obvious that any \hcstp query $q(s, t, k)$ can be obtained by concatenating two \hcsp queries on $G$ and $G_r$,  namely  $P(q(s, t, k)) = P(\mathcal{q}_{s, \ceil{q.k/2}, G}) \oplus P(\mathcal{q}_{t, \floor{q.k/2}, G_r})$.  The benefits of   introducing \hcsp query are twofold: (1) the sharing opportunities among different \hcstp queries are  still remained based on the \hcsp queries; (2) the definition of \hcsp  follows the framework of \kw{PathEnum}, which means it is potentially easy to integrate the \hcsp-based computation sharing strategy into  \kw{PathEnum}. The following problem is to detect the possible common computation based on \hcsp queries. We define:





\begin{definition}
\label{def:hcsubquery}
\textbf{(HC-s Path Query Domination $\prec$)} Given two \hcsp queries $\mathcal{q}_{v, k, G}$ and $\mathcal{q}_{v', k', G}$, $\mathcal{q}_{v', k', G}$ dominates $\mathcal{q}_{v, k, G}$ if $k' \le k - \kw{dist}(v, v')$, denoted by $\mathcal{q}_{v', k', G} \prec \mathcal{q}_{v, k, G}$.
\end{definition}

\vspace{1mm}

We have the following lemmas on \hcsp queries:

\begin{lemma}
\label{lem:enum_pre}
Given two \hcsp queries  $\mathcal{q}_{v, k, G}$ and $\mathcal{q}_{v', k', G}$, if $\mathcal{q}_{v', k', G} \prec \mathcal{q}_{v, k, G}$, then  for any $p_B \in P(\mathcal{q}_{v, k, G})$, there exists $p_A \in P(\mathcal{q}_{v', k', G'})$ such that $p_A \subseteq p_B$. 
\end{lemma}

\myproof According to \refdef{hcsubquery}, because $k - \kw{dist}(v, v') \ge k'$, we have $P(q(v, v', \kw{dist}(v, v'))) \oplus P(\mathcal{q}_{v', k', G}) \subseteq P(\mathcal{q}_{v, k, G})$. This directly implies that for for any $p_B \in P(\mathcal{q}_{v, k, G})$, there exists $p_A \in P(\mathcal{q}_{v', k', G'})$ such that $p_A \subseteq p_B$. $\eop$

\begin{lemma}
\label{lem:enum_share}
Given a \hcstp query $q(s, t, k)$, if there exists two \hcsp queries $\mathcal{q}_{A}$ and $\mathcal{q}_{B}$ on $G$ and $G_r$ respectively s.t. $\mathcal{q}_{A} \prec \mathcal{q}_{s, \ceil{q.k/2}, G}$ and $\mathcal{q}_{B} \prec \mathcal{q}_{t, \floor{q.k/2}, G_r}$, then for any $p' \in P(\mathcal{q}_{A}) \oplus P(\mathcal{q}_{B})$, there exists $p \in P(q)$ such that $p' \subseteq p$.
\end{lemma}

\myproof This lemma can be proved similarly as \reflem{enum_pre}.


%

Based  on \refdef{subquery}, a \hcstp query can be treated as two \hcsp queries $\mathcal{q}_{s, \ceil{q.k/2}, G}$ and $\mathcal{q}_{t, \floor{q.k/2}, G_r}$. Meanwhile, according to \reflem{enum_share}, $\mathcal{q}_{s, \ceil{q.k/2}, G}$/$\mathcal{q}_{t, \floor{q.k/2}, G_r}$ can be further obtained based on the \hcsp queries dominating them. Therefore, for a set of \hcstp enumeration queries $Q$, their common computation can be characterized by the dominating \hcsp queries and the whole \hcstp enumeration cost can be reduced if the dominating \hcsp queries can be efficiently detected and reused. 

Ideally, we hope to find a global optimal set of \hcsp queries such that all the common computation can be shared without any duplicate re-computation. However, due to the NP-hardness of the problem, it is  very likely that the overhead caused by  computing these optimal \hcsp queries significantly outweighs the benefits obtained due to common computation sharing.  Therefore, in the following, we propose a two-phase light-weight method to compute dominating \hcsp queries, which aims to share the common computation among $Q$ as much as possible.

\stitle{\underline{Phase 1: Query clustering.}} The aim of query clustering is to  group the queries that possibly share a large amount of  common computation together to avoid the submarginal common dominating \hcsp query detection. The challenge is that we do not have any detailed information about the \hcstp queries except the source/target vertex and hop constraint, and thus it is hard to quantify the similarity among different \hcstp queries. To address this problem, we observe that for a \hcstp query $q$,  the vertices in the corresponding \hcstps of $q$ must be reachable in $q.k$ hops from $q.s$/$q.t$ on $G$/$G_r$. Following this observation,  for two \hcstp queries $q_A$ and $q_B$, if they share a large amount of common computation regarding \hcstps, then the same set of vertices tend to be explored when conducting \kw{BFS} traversals from the source/target vertex of $q_A$/$q_B$ on $G$/$G_r$,  which motivates us to define the similarity between queries based on the hop-constrained neighbors as follows:

\begin{definition}
\textbf{(Hop-Constrained Neighbors)} For an \hcstp  query $q(s, t, k)$, $\Gamma(q)$ represents the set of vertices that are reachable in $q.k$ hops from $q.s$ in $G$, $\Gamma_r(q)$ represents the set of vertices that are reachable in $q.k$ hops from $q.t$ in $G_r$. The hop-constrained neighbors of $q$ are the vertices in $\Gamma(q)\cup \Gamma_r(q)$. 
\label{def:encounter}
\end{definition}


Note that for any \hcstp query $q \in Q$, we do not need to compute $\Gamma(q)$ and $\Gamma_r(q)$ specialized for query clustering as these vertices have been explored during the procedure of the index construction and we can reuse the result. Following \refdef{encounter}, we further define the similarity between two \hcstp queries as follows:

\begin{definition}
\label{def:factor}
\textbf{(\hcstp Query Similarity)} Given two \hcstp queries $q_A$ and $q_B$, the similarity between $q_A$ and $q_B$ is defined as $\mu(q_A, q_B) = 2/(\frac{\min(|\Gamma(q_A)|, |\Gamma(q_B)|)}{|\Gamma(q_A) \cap \Gamma(q_B)|}  + \frac{\min(|\Gamma_r(q_A)|, |\Gamma_r(q_B)|)}{|\Gamma_r(q_A) \cap \Gamma_r(q_B)|})$ \footnote{{if $|\Gamma(q_A) \cap \Gamma(q_B)|$ and $|\Gamma_r(q_A) \cap \Gamma_r(q_B)|$ are $0$, $\mu = 0$; If one of $|\Gamma(q_A) \cap \Gamma(q_B)|$ or $|\Gamma_r(q_A) \cap \Gamma_r(q_B)|$ is 0,  the corresponding part $\frac{\min(|\Gamma(q_A)|, |\Gamma(q_B)|)}{|\Gamma(q_A) \cap \Gamma(q_B)|}$ or $\frac{\min(|\Gamma_r(q_A)|, |\Gamma_r(q_B)|)}{|\Gamma_r(q_A) \cap \Gamma_r(q_B)|}|$ is $0$}}. 
\end{definition}

\vspace{1mm}

{ Clearly, \hcstp query similarity has the following properties: (1) $0 \le \mu(q_A, q_B) \le 1$; (2) if $P(q_A)$ is a subset of $P(q_B)$, namely $\forall p \in P(q_A)$, $\exists~ p' \in P(q_B)$ s.t. $p \subseteq p'$, $\mu(q_A, q_B) = 1$; (3) if $P(q_A)$ and $P(q_B)$ have no overlaps, $\mu(q_A, q_B) = 0$. It is worth noting that \refdef{factor} can well characterize the similarity between two \hcstp queries even if they have different $k$ values. For example, if $q_A$ has a much larger $k$ than $q_B$ and $q_B.s$ and $q_B.t$ are on the HC-s-t paths of $q_A$. In this case, $q_B$ shares a large amount of common sub-structures with $q_A$ clearly, and the corresponding  \hcstp query similarity is large as well.}

Based on the \hcstp query similarity, we cluster the queries $Q$ into different groups following  a hierarchical clustering manner \cite{Jo67} as the number of queries in $Q$ is medium in size, and  hierarchical clustering does not require a pre-specified number of clusters and it is easy to implement. When merging two groups during hierarchical clustering, we use the following metric:  

\begin{definition}
\label{def:factor}
\textbf{(Group Similarity)} Given two groups of \hcstp  queries $C_A$ and $C_B$,  the similarity between $C_A$ and $C_B$ is defined as $\delta(C_A, C_B) = \frac{1}{|C_A||C_B|} \cdot \sum_{q_A \in C_A, q_B \in C_B} {\mu(q_A, q_B)}.$
\end{definition}

\SetKwFor{Loop}{Loop}{}{EndLoop}
\begin{algorithm}[!tb]
  \SetAlgoLined\DontPrintSemicolon

  \nl $C \leftarrow$ map $Q$ into $|Q|$ clusters, each containing a query $q \in Q$; \\
  \nl \While{changes have been made to $C$}{
  \nl $\kw{sim} \leftarrow 0$; \\ 
  \nl \For{$i \in 0..|C|$}{
  \nl \For{$j \in (i+1)..|C|$}{
  \nl \If{$\delta(C[i], C[j]) > \kw{sim}$}{
  \nl $\kw{sim} \leftarrow \delta(C[i], C[j])$; $C_0 \leftarrow C[i]; C_1 \leftarrow C[j]$;\\
  }}}

  \nl \If{$\kw{sim} > \gamma$}{
  \nl merge $C_0$ and $C_1$ in $C$;
  } 
  }
  \nl \textbf{return} $C$;

  \caption{\kwnospace{ClusterQuery}($Q, \gamma$)}
  \label{alg:cluster}
\end{algorithm}



\stitle{Algorithm.} Our clustering algorithm \querycluster is detailed in \refalg{cluster}. The given parameter $\gamma$ is a threshold to determine when to finish the clustering. Initially, each query in $Q$ represents a group (line 1). Then, \refalg{cluster} iterates all pairs of groups in $C$ and retrieves two groups $C_0$ and $C_1$ with the maximum similarity \kw{sim} in the current clustering (line 4-7). If \kw{sim} is larger than the given similarity threshold $\gamma$, $C_0$ and $C_1$ are merged (line 8-9). The clustering finishes when no two groups exist in $C$ such that their similarity is larger than $\gamma$ (line 2) and the result is returned in line 10.


\begin{figure}[htb]
  \centering
  \includegraphics[width=\columnwidth]{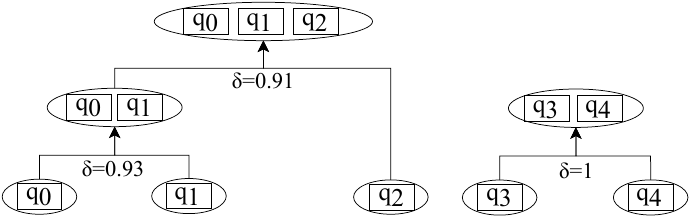}
    \vspace{-2mm}
  \caption{Example of query clustering ($\gamma = 0.8$)}
  \label{fig:cluster}
    \vspace{-2mm}
\end{figure}

%

\begin{example}
Reconsider graph $G$ and the \hcstps queries $Q$ in \reffig{graph}. The clustering procedure in \querycluster with $\gamma = 0.8$ is demonstrated in \reffig{cluster}. Initially, the query similarity between every pair of queries is computed. For instance, as $\Gamma(q_3) = \{v_4, v_9, v_3,  v_8, v_{15}, v_6,  v_{11}, v_{13}, v_{14}\}$ and $\Gamma(q_4) = \{v_9, v_3, v_8, v_{15}, v_6, v_{11}, v_{13}, v_{14}\}$,  $\frac{\min(|\Gamma(q_3)|, |\Gamma(q_4)|)}{|\Gamma(q_3) \cap \Gamma(q_4)|}$ is calculated as 1. Similarly, $\frac{\min(|\Gamma_r(q_3)|, |\Gamma_r(q_4)|)}{|\Gamma_r(q_3) \cap \Gamma_r(q_4)|}  = 1$, which results in $\mu(q_3, q_4) = 1$. As the query similarity between $\{q_3\}$ and $\{q_4\}$ is the largest of all query cluster pairs, they are immediately grouped together. $\{q_0\}$ and $\{q_1\}$ are then chosen in the same way as they have the second largest similarity $\delta(\{q_1\}, \{q_2\}) = 0.93$, and hence grouped together. Then, because $\delta(\{q_2\}, \{q_0, q_1\})=0.91 > \delta(\{q_2\}, \{q_3, q_4\})=0$, $\{q_2\}$ is merged into the cluster $\{q_0, q_1\}$. Subsequently, as it is found that $\delta(\{q_0, q_1, q_2\}, \{q_3, q_4\}) = 0.64 < \gamma$, \querycluster terminates with two groups of queries: $\{q_0, q_1, q_2\}$ and $\{q_3, q_4\}$.
\end{example}

\begin{figure}[!htb]
 \begin{subfigure}[b]{0.5\columnwidth}
  \centering
  \includegraphics[width=\columnwidth]{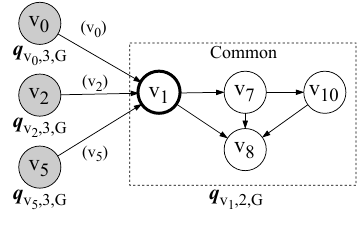}
      \vspace{-6mm}
  \caption{}
  \end{subfigure}
   \begin{subfigure}[b]{0.48\columnwidth}
  \centering
  \includegraphics[width=\columnwidth]{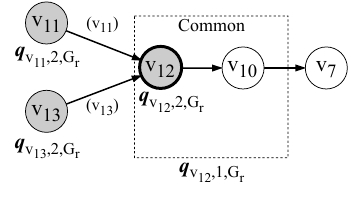}
      \vspace{-6mm}
  \caption{}
  \end{subfigure}
     \vspace{0mm}
     \caption{Example of common \hcsp query}
  \label{fig:icommon}
      \vspace{-2mm}
\end{figure}



\begin{algorithm}[htb]
  \SetAlgoLined\DontPrintSemicolon
  
  \nl $M_E \leftarrow \emptyset $; $M_Q \leftarrow \emptyset $; $\Psi \leftarrow$ empty query sharing graph;\\
  \nl \ForEach{$q \in Q$} {
  \nl $\Psi.\kw{addEdge}((\mathcal{q}_{q.s, \ceil{q.k/2}, G}, q))$;\\
  \nl $M_E[q.s].\kw{add}(\mathcal{q}_{q.s, \ceil{q.k/2}, G})$;
  }
   \nl $k_{max} \leftarrow \max_{q \in Q}{\ceil{q.k/2}}$; \\
  \nl \ForEach{$k \in 0..k_{max}$}{
  \nl \ForEach{$v \in V(G)$}{
  \nl $S_Q \leftarrow \emptyset$; \\
  \nl \ForEach{$\mathcal{q} \in M_E[v]$}{
  \nl \If{$\mathcal{q}.k = k_{max} - k$}{
  \nl $S_Q.\kw{add}(\mathcal{q})$; $M_E[v].\kw{remove}(\mathcal{q})$;
  }}
  \nl \If{$S_Q = \emptyset$ } {
  \nl \textbf{continue};
  }
  \nl \ElseIf{$|S_Q| = 1$}{
  \nl $M_Q[v] \leftarrow S_Q[0]$;
  }
  \nl \Else{
  \nl \ForEach{$\mathcal{q} \in S_Q$}{
  \nl $\Psi.\kw{addEdge}((\mathcal{q}_{v, k_{max}-k, G}, \mathcal{q}))$;
  }
  \nl $M_Q[v] \leftarrow \mathcal{q}_{v, k_{max}-k, G}$;
  }
  \nl \ForEach{$v' \in \ \nbrout(v)$ s.t. $v'$ meets the hop constraint}{
  \nl \If{$M_Q[v'] \ne \emptyset$ and $M_Q[v] \not\prec M_Q[v']$}{
  \nl $\Psi.\kw{addEdge}((M_Q[v'], M_Q[v]))$;
  }
  \nl \Else{
  \nl $M_E[v'].\kw{add}(M_Q[v])$;
  }
  }
  }
  }
  
  \nl \textbf{return} $\Psi$;
  
  \caption{\overlapextract($G, Q$)}
  \label{alg:identify}
\end{algorithm}

\begin{figure*}[!htb]
 \begin{subfigure}[b]{1\columnwidth}
  \centering
  \includegraphics[width=\columnwidth]{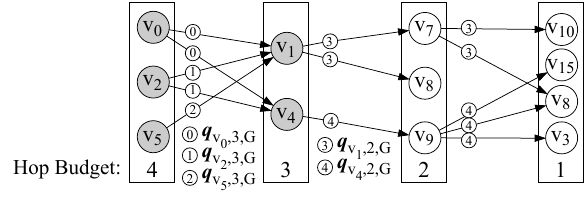}
      \vspace{-8mm}
  \caption{Detection process on $G$}
  \end{subfigure}
  \hspace{0.9cm}
   \begin{subfigure}[b]{0.85\columnwidth}
  \centering
  \includegraphics[width=\columnwidth]{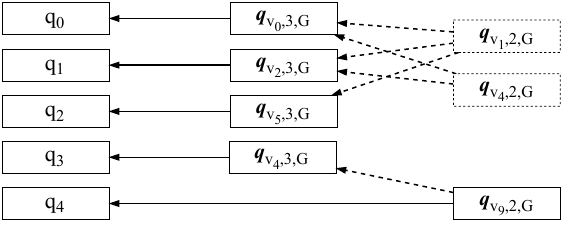}
      \vspace{-8mm}
  \caption{Query sharing graph $\Psi$}
  \end{subfigure}
     \vspace{-1mm}
     \caption{Common \hcsp query detection}
  \label{fig:identify}
      \vspace{-5mm}
\end{figure*}

\stitle{\underline{Phase 2: Common  \hcsp queries detection.}} After the clustering step, the queries in the same group tend to share a large amount of common computation.  The next step is to detect such common computation and reuse them to speed up the processing. Due to the NP-hardness of the  problem, we focus on the common computation among the \hcstp queries characterized by their dominating \hcsp queries, and  propose a light-weight dominating \hcsp queries detection algorithm based on the following observations:

Consider $G$ and queries $Q$ in \reffig{graph}, $q_0$, $q_1$, and $q_2$ are clustered in the same group in step 1. Recall that the results of a \hcstp query  can be obtained by the concatenation of two \hcsp queries on $G$ and $G_r$.  \reffig{icommon} (a) shows the enumeration procedures of the \hcsp queries $\mathcal{q}_{v_0, 3, G}$, $\mathcal{q}_{v_2, 3, G}$  and $\mathcal{q}_{v_5, 3, G}$, which correspond to  $q_0$, $q_1$ and $q_2$ on $G$. The source vertices of these initial queries are marked in grey. During the enumeration, after consuming one hop of their hop budgets (the initial hop budget is the same as hop constraint, and the remaining hop budget decreases as the enumeration proceeds), all three \hcsp queries encounter the same vertex $v_1$, and they will enumerate the same paths in the following enumeration steps, namely  $\{(v_1, v_7, v_{10}), (v_1, v_7, v_{8}), (v_1, v_8)\}$, which can be represented by the dominating  \hcsp query $\mathcal{q}_{v_1, 2, G}$ of $\mathcal{q}_{v_0, 3, G}$, $\mathcal{q}_{v_2, 3, G}$  and $\mathcal{q}_{v_5, 3, G}$. This inspires us with the idea that when queries encounter the same vertex $v$ with the same hop budget left, the following enumeration represented by the dominating \hcsp query $\mathcal{q}_{v, k, G}$ of these queries is common.  

The above observation shows that the computation with the same remaining hop budget during enumeration can be possibly shared by identifying the corresponding  dominating \hcsp queries. Furthermore, the identified dominating \hcsp queries still possibly contain common computation.   Consider $\mathcal{q}_{v_{11}, 2, G_r}$, $\mathcal{q}_{v_{13}, 2, G_r}$ and $\mathcal{q}_{v_{12}, 2, G_r}$ in \reffig{icommon} (b), which correspond to   the \hcsp queries of $q_0$, $q_1$ and $q_2$ on $G_r$. For $\mathcal{q}_{v_{11}, 2, G_r}$ and $\mathcal{q}_{v_{13}, 2, G_r}$, after consuming one hop of their hop budgets, both of them encounter vertex $v_{12}$, and the enumeration from $v_{12}$ with a remaining hop budget of 1, represented by the dominating \hcsp query $\mathcal{q}_{v_{12}, 1, G_r}$, can be shared between them. Meanwhile, $\mathcal{q}_{v_{12}, 2, G_r}$ which corresponds to $q_2$ already exists. As $\mathcal{q}_{v_{12}, 2, G_r}$ has a greater hop budget than $\mathcal{q}_{v_{12}, 1, G_r}$, the results of $\mathcal{q}_{v_{12}, 1, G_r}$ can be directly obtained during the enumeration of $\mathcal{q}_{v_{12}, 2, G_r}$, which implies that we can further speed up the processing by  sharing computation among the identified dominating \hcsp queries with the same source vertex but different hop budget $k$.

Following these two observations, we devise our algorithm to detect the dominating \hcsp queries. Before introducing the detailed algorithm, we first define the query sharing graph, which records the computation sharing relations among \hcstp/\hcsp queries  and facilitates the computation sharing during the final enumeration.


\begin{definition}
\label{def:query_graph}
\textbf{(Query Sharing Graph $\Psi$)}{ The query sharing graph $\Psi$ is a directed graph where a node $v$ represents either a \hcstp query or a \hcsp query, and  an edge $(u, v)$ indicates that the \hcsp query $\mathcal{q}_{v, k, G}$ represented by $v$  dominates the \hcsp query $\mathcal{q}_{u, k', G}$ represented by $u$ or  the generation relationship between \hcstp query $q(s, t, k)$ represented by $u$ and \hcsp query $\mathcal{q}_{s, \ceil{q.k/2}, G} / \mathcal{q}_{t, \floor{q.k/2}, G_r}$ represented by $v$.}
\end{definition}

\stitle{Algorithm.} Based on the query sharing graph $\Psi$,  our dominating \hcsp queries detection algorithm is shown in \refalg{identify}. \refalg{identify} only shows the case on $G$, and the dominating \hcsp queries  on  $G_r$ can be detected in the same way.  Given a graph $G$ and a batch of queries $Q$, \refalg{identify} uses two hash maps $M_E$ and $M_Q$. In $M_E$,  each entry $(v, Q')$ stores the set of queries $Q'$ that are extending vertex $v$ at the current size of the remaining hop budget. In $M_Q$, each entry $(v, \mathcal{q}_{v, k, G})$  records the latest \hcsp identified on vertex $v$ (line 1).  For $q \in Q$, it adds a trivial edge $(\mathcal{q}_{q.s, \ceil{q.k/2}, G}, q)$ to $\Psi$ and $\mathcal{q}_{q.s, \ceil{q.k/2}, G}$  to $M_{E}[q.s]$ indicating its prefix is initially extending from $q.s$ (lines 2-4). 

After this, \refalg{identify} first computes the maximum size of hop budgets among all queries as $k_{max}$ (line 5). Then, it iterates through each possible size of the remaining hop budgets and detects the vertices where the common dominating \hcsp queries may exist (lines 6-24). For each size of the consumed hops  $k$ and for each vertex $v$, a set $S_Q$ is created to store the \hcsp queries that have the same hop budget of $k_{max} - k$ (lines 7-8). Then, the queries extending $v$ in $M_E[v]$ that have a hop budget of $k_{max} - k$ are added to $S_Q$ (lines 9-11). If $S_Q$ is empty, vertex $v$ is not extended by any \hcsp query and can be skipped (lines 12-13). Else, if there is only one query in $S_Q$, this query is set as the most recent \hcsp that starts from $v$ (lines 14-15). Otherwise, it means that $v$ is extended by multiple queries with the same remaining hop budget, thus a new dominating \hcsp query $\mathcal{q}_{v, k_{max}-k, G}$ is identified and the edges from it to the queries in $S_Q$ are added to $\Psi$ (lines 16-19). After identifying the dominating \hcsp queries between the queries with the same remaining hop budget, \refalg{identify} iterates through the out-neighbors of $v$ and checks if their results can be derived from those of existing identified \hcsp queries (lines 20-24). Specifically, for each out-neighbor $v'$ of $v$ that meets the hop constraint, if there is an existing \hcsp query $M_Q[v']$ starting from $v'$ and $M_Q[v]$ is not a dominating \hcsp query of $M_Q[v']$, then a new edge from $M_Q[v']$ to $M_Q[v]$ is inserted into $\Psi$ (lines 20-22). Otherwise, $M_Q[v]$ is added to $M_E[v']$, indicating that its prefix is going to subsequently extend $v'$ (lines 23-24). Finally, $\Psi$ is returned (line 25).




\begin{example}
Reconsider graph $G$ and the \hcstp queries $Q$ in \reffig{graph}. After running the clustering procedure with $\gamma = 0.8$, the queries are divided into two groups: $C_0 = \{q_0, q_1, q_2\}$ and $C_1 = \{q_3, q_4\}$, shown in \reffig{cluster}. We demonstrate the example of running \refalg{identify} on $(G, C_0)$ in detail, where the vertices extended in each iteration with each size of the remaining hop budget are shown in \reffig{identify} (a). The vertices on which a new \hcsp query is identified are marked in grey. Additionally, \reffig{identify} (b) demonstrates the changes on the vertices and edges of the query sharing graph $\Psi$ when identifying common \hcsp query on both $C_0$ and $C_1$. The solid vertices and edges are those inserted when $\Psi$ is initialized, while the dashed vertices and edges are those inserted during the detection procedure. 

Initially, the \hcsp query of each \hcstp query in $C_0$ on $G$ is extracted, resulting in the \hcsp queries $\{\mathcal{q}_{v_0, 3, G}, \mathcal{q}_{v_2, 3, G},\mathcal{q}_{v_5, 3, G}\}$. They are subsequently inserted into $\Psi$ shown in \reffig{identify} (b). Obviously, the maximum hop budget $k_{max}$ among all identified \hcsp queries is $3$. In the first iteration, because $\mathcal{q}_{v_0, 3, G}$, $\mathcal{q}_{v_2, 3, G}$ and $\mathcal{q}_{v_5, 3, G}$ all have a hop budget of 3, their out-neighbors to be extended are examined. For vertex $v_1$, as it is extended by all three queries, then $M_E[v_1] = \{\mathcal{q}_{v_0, 3, G}, \mathcal{q}_{v_2, 3, G},\mathcal{q}_{v_5, 3, G}\}$. As a result, the three queries $\mathcal{q}_{v_0, 3, G}$, $\mathcal{q}_{v_2, 3, G}$ and $\mathcal{q}_{v_5, 3, G}$ in $M_E[v_1]$ are converted into a dominating \hcsp query $\mathcal{q}_{v_1, 2, G}$, which is inserted into $\Psi$ with the three sharing queries as its out-neighbors. $M_Q[v_1]$ is also set to $\mathcal{q}_{v_1, 2, G}$. Similarly, as $v_4$ is extended by $\mathcal{q}_{v_0, 3, G}$ and $\mathcal{q}_{v_2, 3, G}$, $M_E[v_4] = \{\mathcal{q}_{v_0, 3, G}, \mathcal{q}_{v_2, 3, G}\}$ and these two queries are converted into a dominating \hcsp query $\mathcal{q}_{v_4, 2, G}$. Then, $\mathcal{q}_{v_4, 2, G}$ is inserted into $\Psi$ with $\mathcal{q}_{v_0, 3, G}$ and $\mathcal{q}_{v_2, 3, G}$ as its out-neighbors. $M_Q[v_4]$ is subsequently set to $\mathcal{q}_{v_4, 2, G}$. After this, the newly identified dominating \hcsp query $\mathcal{q}_{v_1, 2, G}$ extends $v_7$ and $v_8$, and $\mathcal{q}_{v_4, 2, G}$ extends $v_9$. This results in $M_E[v_7] = M_E[v_8] = \{\mathcal{q}_{v_1, 2, G}\}$ and $M_E[v_9] = \{\mathcal{q}_{v_4, 2, G}\}$. Because no vertex is commonly visited by multiple \hcsp queries, no new dominating \hcsp query is detected. In the last iteration,  $\{v_{10}, v_8\}$ are extended by $\mathcal{q}_{v_1, 2, G}$ and $\{v_{15}, v_8, v_3\}$ are extended by $\mathcal{q}_{v_4, 2, G}$. As the remaining hop budget has dropped to 0, the identification process of $C_0$ on $G$ terminates. 
\end{example}

\begin{theorem}
Given a batch of queries $Q$ and a graph $G$, \refalg{identify} has a time complexity of $O(|Q|(|V(G)| + |E(G)|))$ and a space complexity of $O(|V(G)|^2)$.
\end{theorem}

\myproof Based on the procedure of \refalg{identify}, each query visits every vertex and edge in $G$ for at most once. Thus the time complexity for \refalg{identify} is $O(|Q|(|V(G)| + |E(G)|))$. Moreover, each unique vertex in $G$ can at most appear once in the query sharing graph $\Psi$. This is because for any vertex $v$, if two \hcsp queries $\mathcal{q}_{v, k, G}$ and $\mathcal{q}_{v, k', G}$ with $k \ge k'$ are identified, the results of $\mathcal{q}_{v, k', G}$ will be derived directly from the results of $\mathcal{q}_{v, k, G}$ and only $\mathcal{q}_{v, k, G}$ will be added to $\Psi$. Hence, $\Psi$ has at most $|V(G)|$ nodes and $|V(G)|^2$ edges, resulting in a space complexity of $O(|V(G)|^2)$. $\eop$

\begin{algorithm}[!htb]
  \SetAlgoLined\DontPrintSemicolon
  \SetKwFunction{proc}{proc}
   \nl $S \leftarrow \bigcup_{q \in Q} \{q.s\}$; $T \leftarrow  \bigcup_{q \in Q} \{q.t\}$\\
  \nl construct index by  multi-source \kwnospace{BFS}s from $S$ and $T$;\\
  \nl $R \leftarrow$ an empty cache; $C_s \leftarrow \querycluster(Q, \gamma)$; \\
  \nl \ForEach{$C \in C_s$}{
  \nl $\Psi / \Psi_r \leftarrow \overlapextract(G/G_r, C);$\\
  \nl \For{$q \in V(\Psi) / V(\Psi_r)\ in\ topological\ order$}{
  \nl \If{$q$ is a \hcsp query}{
  \nl $P \leftarrow \emptyset;$\\
  \nl $\kw{Search}(G/G_r, P, R, (q.v), q, \Psi / \Psi_r);$\\
  \nl $R.\kw{add}(q, P);$
  }
  \nl \If{$q$ is a \hcstp query and its two \hcsp queries $q_l, q_r$ are computed} {
  \nl \ForEach{$p \in R[q_l] \oplus R[q_r]$} {
  \nl \textbf{if} $p$ has no duplicated vertex \textbf{then} \kwnospace{Output} $p$;
  }}
  \nl \ForEach{$q' \in \Psi/\Psi_r.\nbrin(q)$}{
  \nl \If{$q'' \in \Psi/\Psi_r.\nbrout(q')\ all\ processed$} {
  \nl $R.\kw{remove}(q')$;
  }
  }
  
  }
  
  }
  
  \nl \SetKwProg{myproc}{Procedure}{}{}
  \myproc{\kw{Search}($G, P, R, p, q, \Psi$)}{
  \nl $v' \leftarrow p[|p|]$; $P.\kw{add}(p)$;\\
  \nl \textbf{if} $|p| = q.k$ \textbf{then}  \KwRet ; \\
  \nl \ForEach{$v'' \in G.\nbrout(v')$  s.t. $v''$ meets the hop constraint} {
  \nl \textbf{if} $v'' \in p$ \textbf{then continue}; \\ 
  
  \nl \If{$\exists q' \in \Psi.\nbrin(q)$ s.t. $q'.v = v''$} {
  \nl $P.\kw{add}(p \oplus R[q'])$;
  } 
  \nl \textbf{else} $\kw{search}(G, P, R ,p \bigcup \{v''\}, q, \Psi)$;
  
  }}
  
  \caption{\batchenum($G, Q, \gamma$)}
  \label{alg:overview}
\end{algorithm} 

\vspace{-1mm}

\subsection{\kwnospace{HC}\textrm{-}\kw{s} \kw{Path} Shared Enumeration}
\label{sec:bidirectional}

After detecting the common \hcsp queries in $\Psi$, we can materialize the detected common \hcsp queries and accelerate the batch \hcstp query processing by reusing the materialized results. Based on the procedure of \refalg{identify}, it is obvious that $\Psi$ is a directed acyclic graph.  To maximize the common computation sharing, the queries in $\Psi$ are processed in a topological order and a query $q$ is not processed until all its in-neighbors have been processed. For computation reusage, the common \hcsp queries are computed and stored in a cache $R$.   The detailed algorithm, \batchenum, is shown in \refalg{overview}.

\stitle{Algorithm.} For the batch of queries $Q$, \batchenum first performs two multi-source BFSs from $S$ and $T$, respectively, to compute $\kw{dist}_G(s, v)$ and $\kw{dist}_{G_r}(t, v)$ for all $v \in V(G)$, where $s \in \cup_{q \in Q} \{q.s\}$, $t \in \cup_{q \in Q} \{q.t\}$ (line 1-2). \batchenum then initializes an empty cache $R$ to store the results for the detected dominating \hcsp queries and clusters the queries in $Q$ into different groups (lines 3). In every query group, it identifies the dominating \hcsp queries and stores them in the query sharing graphs $\Psi$ and $\Psi_r$ on $G$ and $G_r$, respectively (lines 4-5). After this, for each query $q \in \Psi$/$\Psi_r$ in the topological order, \batchenum conducts the enumeration on $G/G_r$ if $q$ is a \hcsp query. The paths obtained by the search are stored in the cache (lines 6-10). Otherwise, if the \hcsp queries of the \hcstp query $q$ in both directions have been solved, the paths of the \hcsp queries in both directions are concatenated by $\oplus$ (lines 11-12), and if no duplicated vertex exists in the concatenated path, it is a \hcstp and \batchenum outputs the path (line 13). The results of $q$'s in-neighbor query $q'$ in $\Psi$ are removed from the cache if all out-neighbor queries of $q'$ in $\Psi$ have been processed (lines 14-16). Procedure \kw{Search} enumerates the \kwnospace{HC}-\kwnospace{s} \kw{paths} with hop constraint $q.k$ based on the index recursively (lines 17-24). Specifically, any path $p$ with length smaller than or equal to $q.k$ is is added into $P$ (lines 18-19). For each out-neighbor $v''$ of $v'$ that meets the hop constraint and has not been explored, if there exists a dominating \hcsp query $q'$ where $q'.v = v''$, the \kwnospace{HC}-\kwnospace{s} \kw{paths} are achieved by concatenating $p$ and the cached results for $q'$. (lines 20-23). Otherwise, $v''$ is added in $p$ and the search continues (line 24).

\begin{example}
Reconsider graph $G$ in \reffig{graph} and the query sharing graph constructed in \reffig{identify} (b). The \hcsp queries regarding cluster $\{q_0, q_1, q_2\}$ in $\Psi$ are processed in a topological order of $\{\mathcal{q}_{v_4, 2, G}, \mathcal{q}_{v_1, 2, G}, \mathcal{q}_{v_0, 3, G}, \mathcal{q}_{v_2, 3, G}, \mathcal{q}_{v_5, 3, G}\}$. Initially, the paths for $\mathcal{q}_{v_4, 2, G}$ and $\mathcal{q}_{v_1, 2, G}$ are enumerated as $\{(v_4, v_9, v_{15})$, $(v_4, v_9, v_8)$, $(v_4, v_9, v_3)\}$ and $\{(v_1, v_7, v_{10}), (v_1, v_7, v_8)\}$, respectively. After this, $\mathcal{q}_{v_0, 3, G}$ is processed. As its prefix $(v_0)$ extends both $v_1$ and $v_4$, with its in-neighbor queries $\mathcal{q}_{v_1, 2, G}$ and $\mathcal{q}_{v_4, 2, G}$ computed, its enumeration following $v_1$ and $v_4$ is avoided and its results are directly computed as $\{(v_0, v_1)\} \oplus P(\mathcal{q}_{v_1, 2, G})$ and $\{(v_0, v_4)\} \oplus P(\mathcal{q}_{v_4, 2, G})$, which are consequently $\{(v_0, v_1, v_7, v_{10})$, $(v_0, v_1, v_7, v_8)$,$(v_0, v_4, v_9, v_{15})$, $(v_0, v_4, v_9, v_8)$,$(v_0, v_4, v_9, v_{3})\}$. The results for $\mathcal{q}_{v_2, 3, G}$ and $\mathcal{q}_{v_5, 3, G}$ are enumerated similarly based on concatenating their initial prefixes with the results of their in-neighbor queries. 
\end{example}

\section{Evaluation}
\label{sec:exp}

In this section, we evaluate the efficiency of the proposed algorithms. All the experiments are performed on a machine with one 20-core Intel Xeon CPU E5-2698 and 512 GB main memory running Red Hat Linux 7.3, 64 bit.

\begin{table}[h!]
\caption{Statistics of the datasets}
\centering
\begin{tabular}{c | c | c | c | c | c  } 
 \hline
  Dataset & Name & $|V|$ & $|E|$& $d_{avg}$ & $d_{max}$   \\ 
 \hline
 \kw{Epinsion} & \kw{EP} & 75K & 508K  & 13.4 & 3,079  \\  
 \kw{Slashdot} & \kw{SL} & 82K & 948K & 21.2 & 5,062  \\  
  \kwnospace{Baidu}\textrm{-}\kw{baike} & \kw{BK} & 416K & 3M & 5.0 & 98,173 \\  
 \kw{WikiTalk} & \kw{WT} & 2M & 5M & 5.0 & 1,242 \\  
 \kw{BerkStan} & \kw{BS} & 685K & 7M & 22.2 & 84,290  \\  
  \kw{Skitter} & \kw{SK} & 1.6M & 11M & 13.1 & 35,547  \\  
    \kwnospace{Web}\textrm{-}\kwnospace{uk}\textrm{-}\kw{2005} & \kw{UK} & 130K & 11.7M  & 181.2  & 850 \\ 
  \kwnospace{Rec}\textrm{-}\kw{dating} & \kw{DA} & 169k & 17M & 205.7 & 33,411 \\  
 \kw{Pokec} & \kw{PO} & 1.6M & 31M & 37.5 & 20,518 \\  
 \kw{LiveJournal} & \kw{LJ} & 4M & 69M & 17.9 & 20,333 \\
 \kwnospace{Twitter}\textrm{-}\kw{2010} & \kw{TW} & 42M & 1.46B & 70.5 & 2,997,487 \\
 \kw{Friendster} & \kw{FS} & 65M &  1.81B & 27.5 & 5,214 \\  
 \hline
\end{tabular}
\label{tab:dataset}

\vspace{-3mm}
\end{table}


\begin{figure*}
     \centering
      \includegraphics[width=1.2\columnwidth]{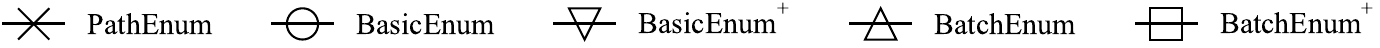}\\
      \includegraphics[width=1\columnwidth]{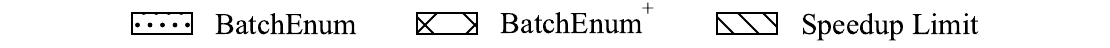}\\

     \begin{subfigure}[b]{0.49\columnwidth}
         \centering
         \includegraphics[width=\columnwidth]{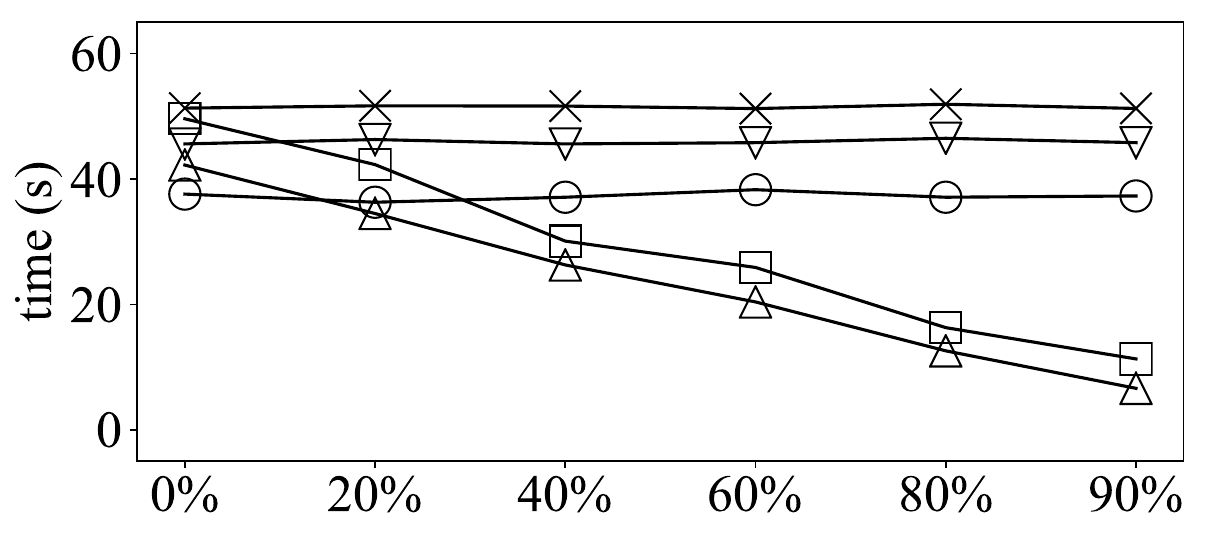}
         \includegraphics[width=\columnwidth]{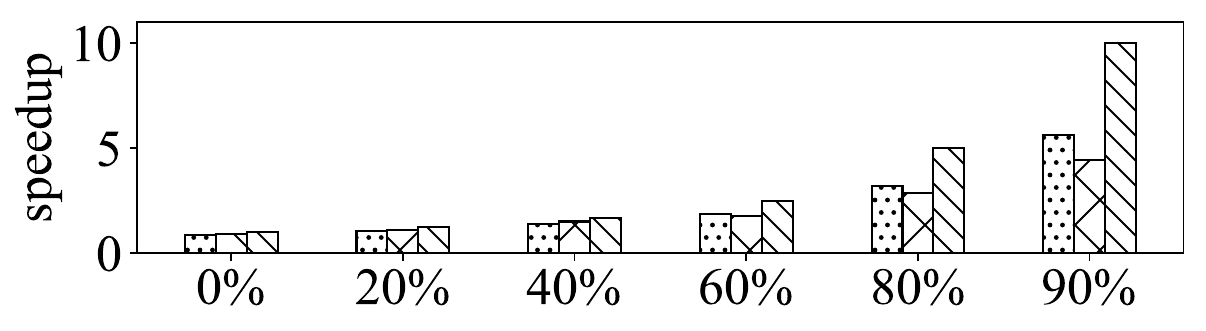}
         \vspace{-6mm}
         \caption{EP}
     \end{subfigure}
     \hfill
     \begin{subfigure}[b]{0.49\columnwidth}
         \centering
         \includegraphics[width=\columnwidth]{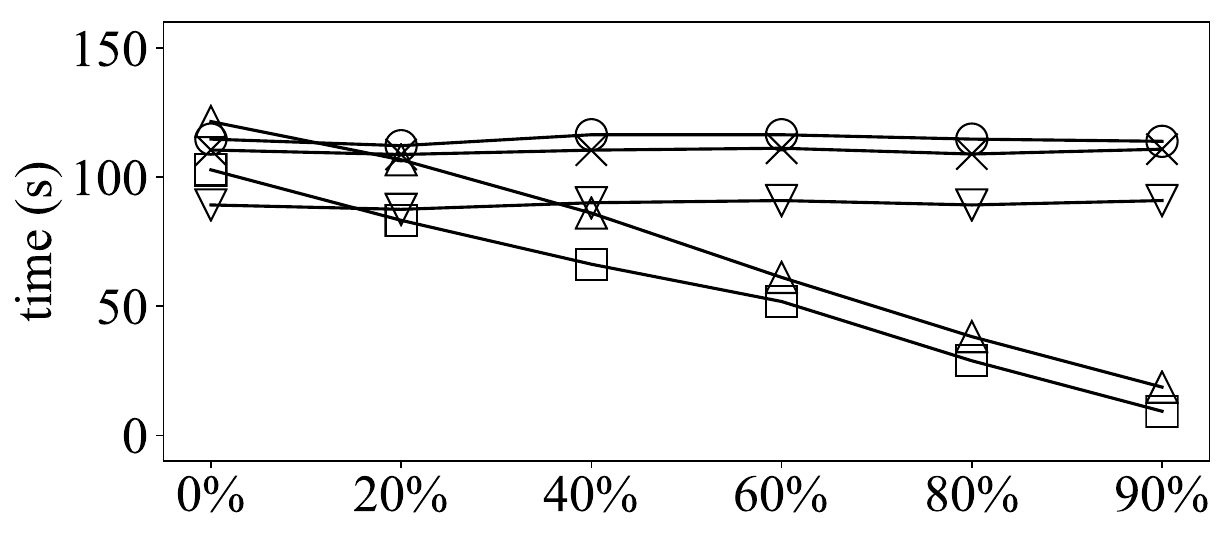}
         \includegraphics[width=\columnwidth]{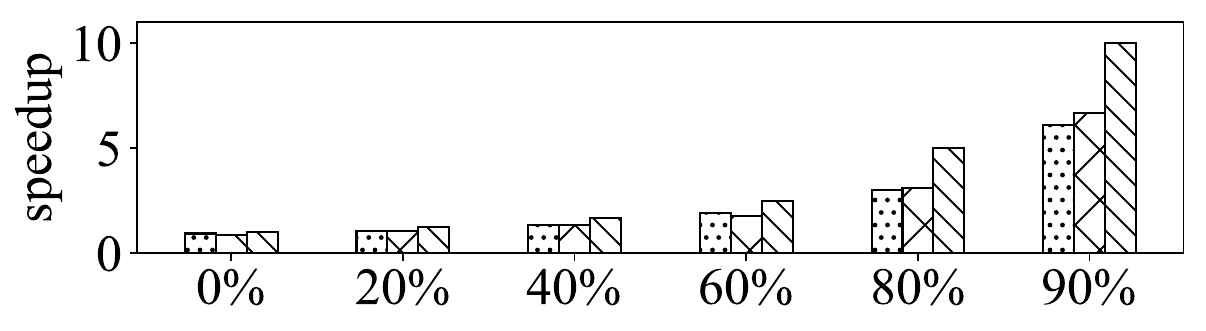}         \vspace{-6mm}
         \caption{SL}
     \end{subfigure}
          \begin{subfigure}[b]{0.49\columnwidth}
         \centering
         \includegraphics[width=\columnwidth]{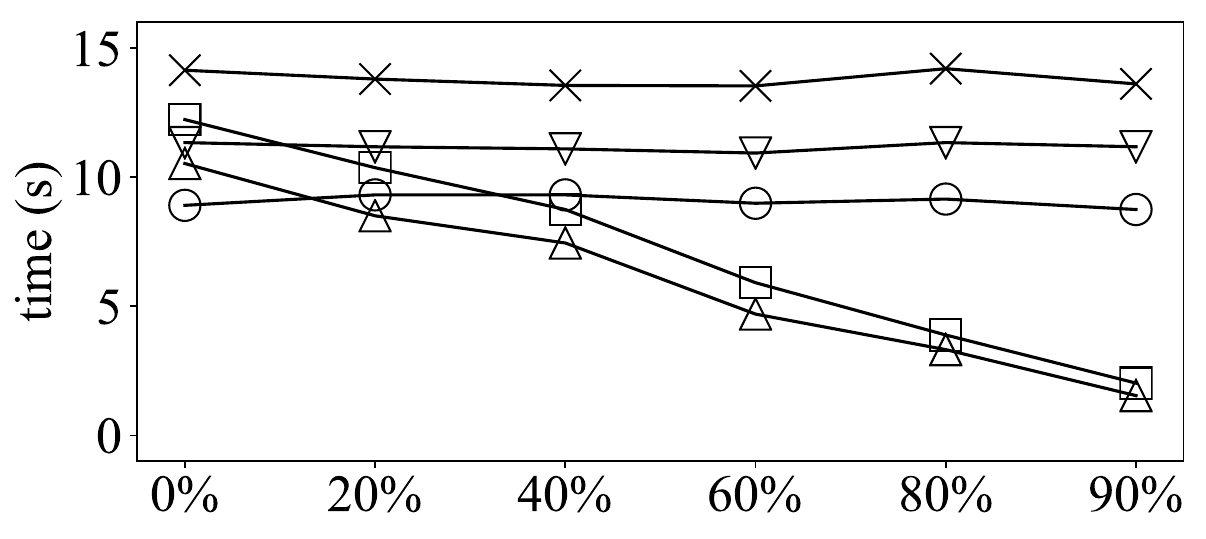}
         \includegraphics[width=\columnwidth]{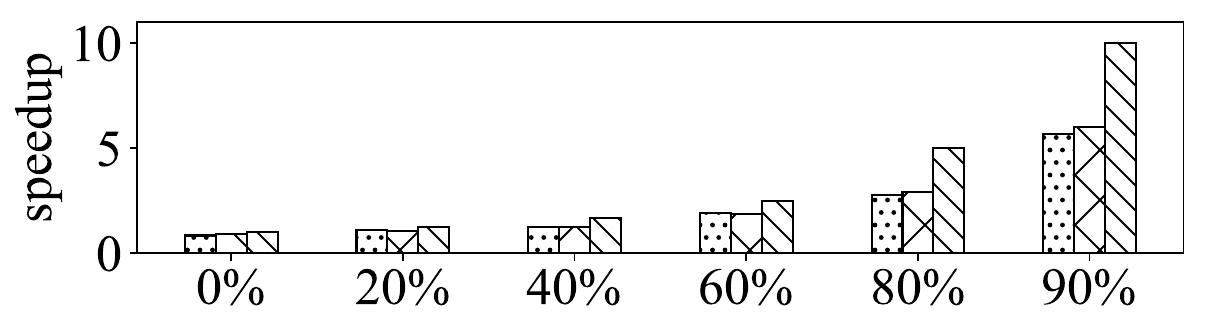}         \vspace{-6mm}
         \caption{BK}
     \end{subfigure}
     \hfill
     \begin{subfigure}[b]{0.49\columnwidth}
         \centering
         \includegraphics[width=\columnwidth]{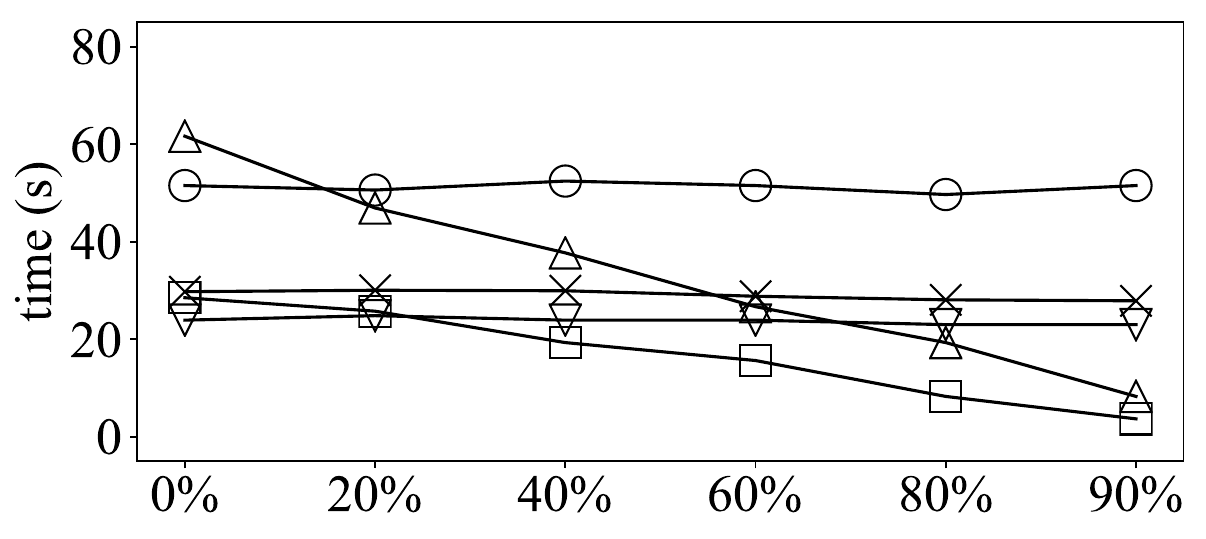}
         \includegraphics[width=\columnwidth]{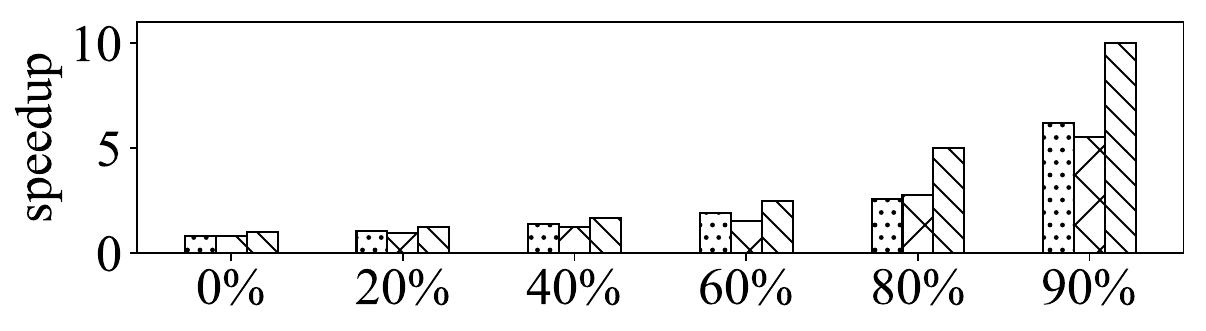}         \vspace{-6mm}
         \caption{WT}
     \end{subfigure}
          \begin{subfigure}[b]{0.49\columnwidth}
         \centering
         \includegraphics[width=\columnwidth]{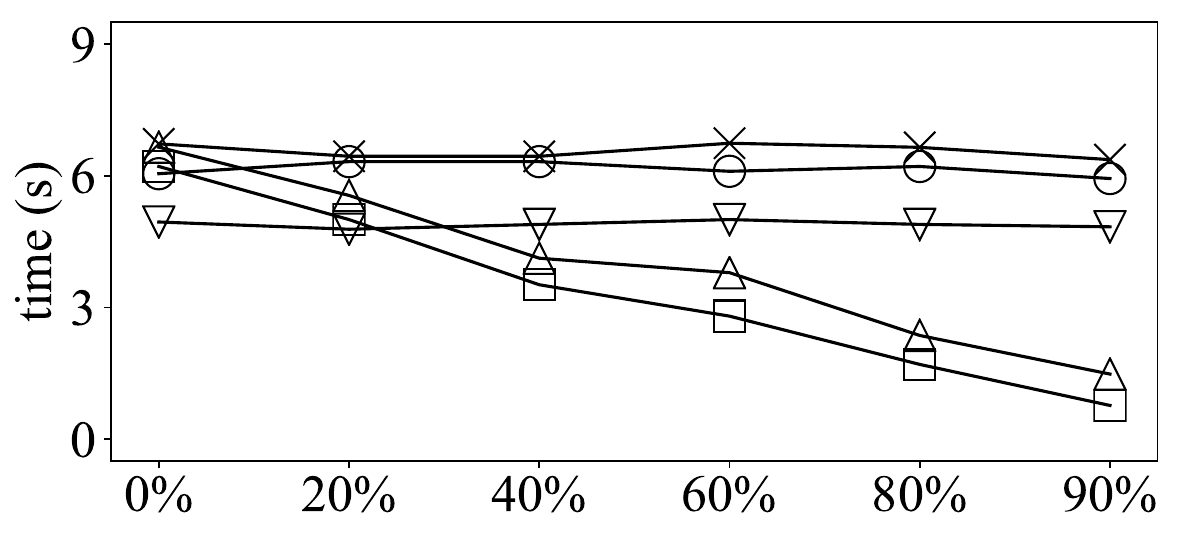}
         \includegraphics[width=\columnwidth]{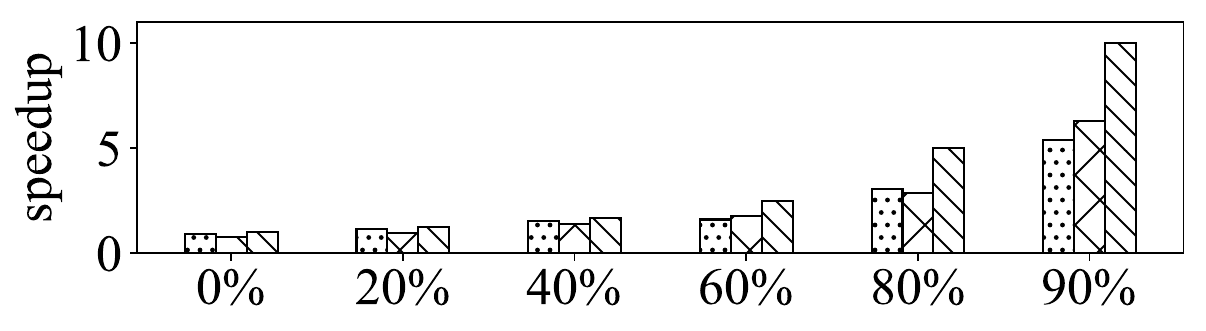}         \vspace{-6mm}
         \caption{BS}
     \end{subfigure}
     \hfill
     \begin{subfigure}[b]{0.49\columnwidth}
         \centering
         \includegraphics[width=\columnwidth]{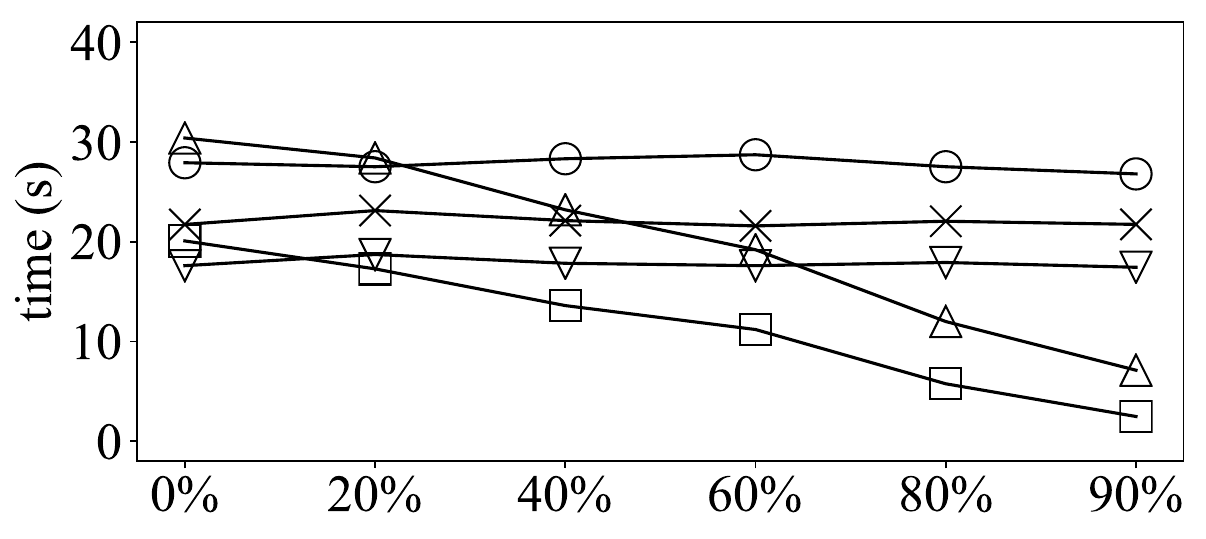}
         \includegraphics[width=\columnwidth]{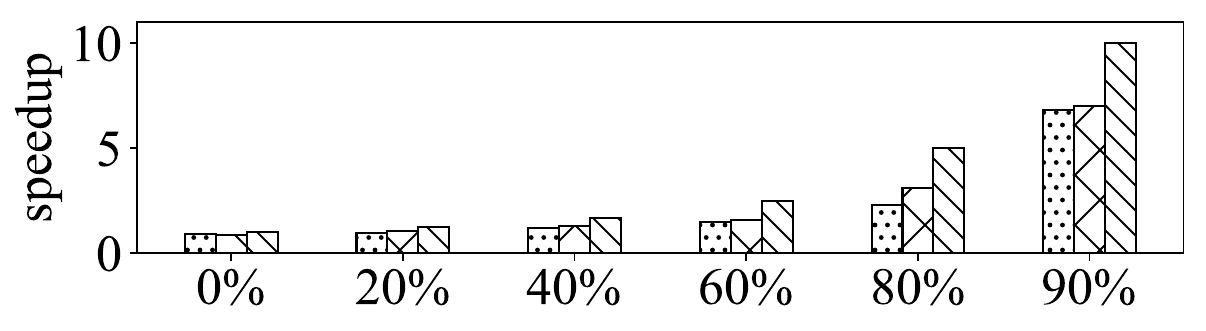}         \vspace{-6mm}
         \caption{SK}
     \end{subfigure}
          \begin{subfigure}[b]{0.49\columnwidth}
         \centering
         \includegraphics[width=\columnwidth]{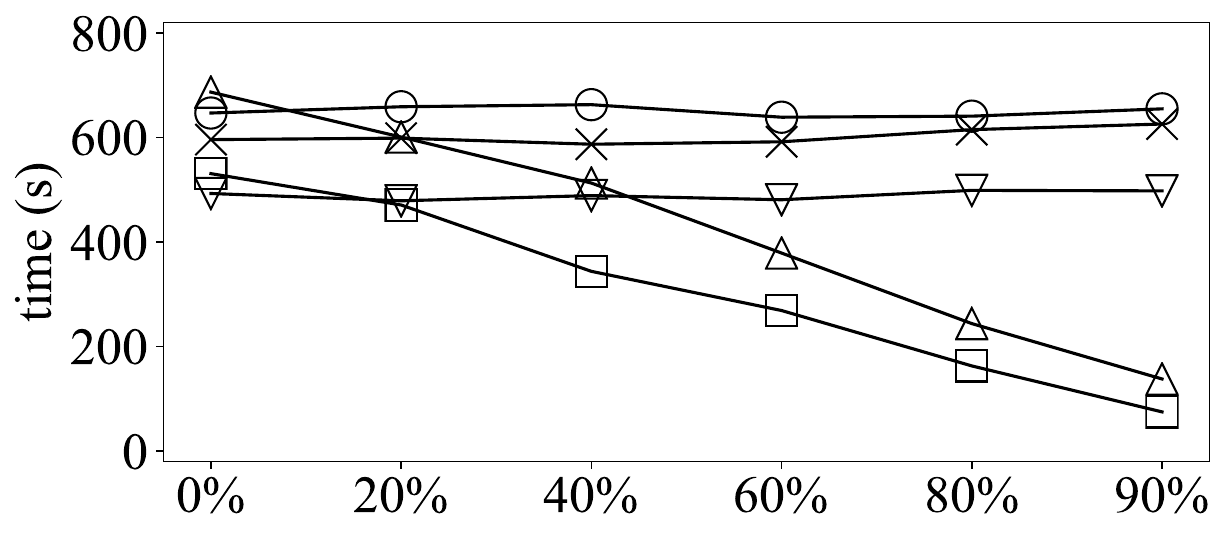}
         \includegraphics[width=\columnwidth]{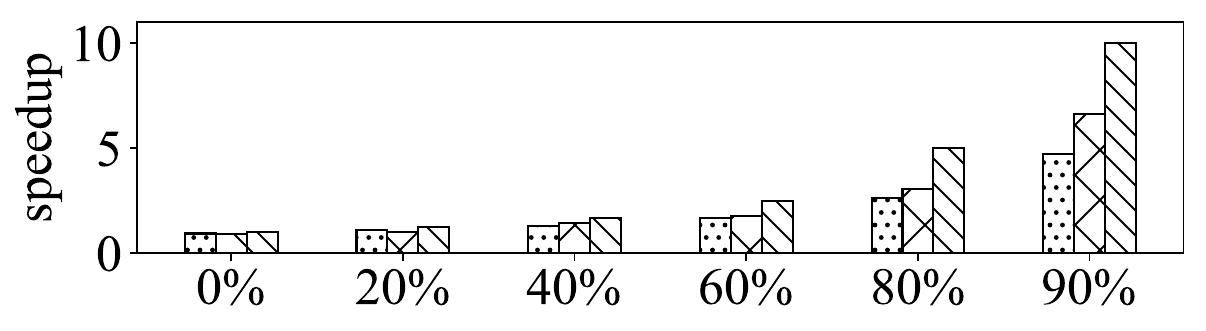}         \vspace{-6mm}
         \caption{UK}
     \end{subfigure}
     \hfill
     \begin{subfigure}[b]{0.49\columnwidth}
         \centering
         \includegraphics[width=\columnwidth]{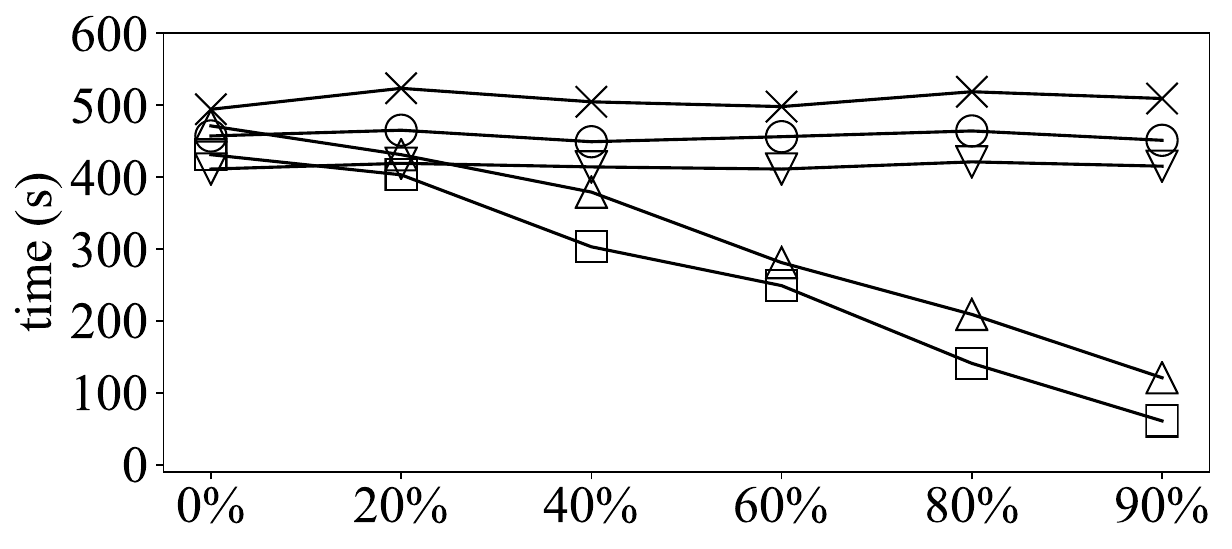}
         \includegraphics[width=\columnwidth]{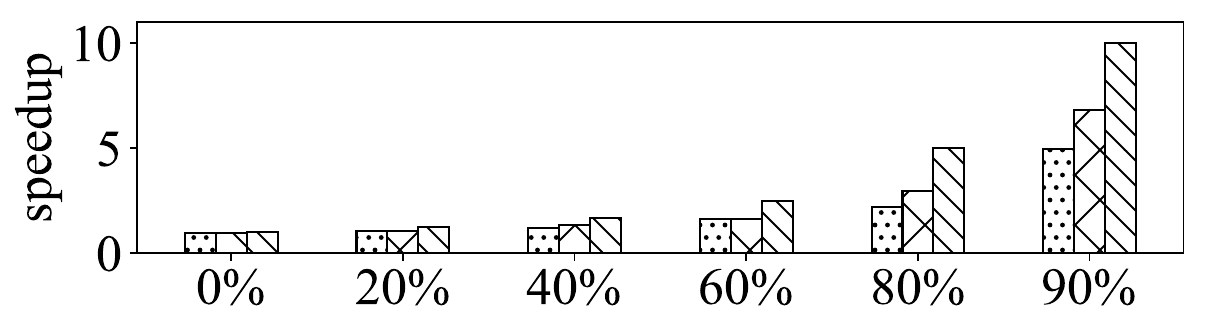}         \vspace{-6mm}
         \caption{DA}
     \end{subfigure}
    \begin{subfigure}[b]{0.49\columnwidth}
         \centering
         \includegraphics[width=\columnwidth]{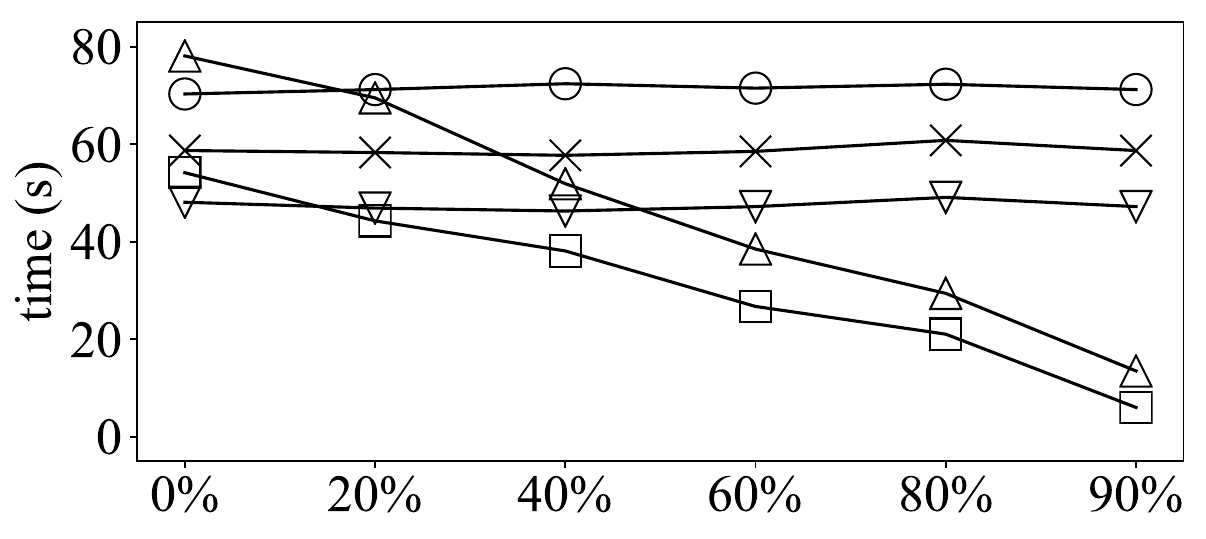}
         \includegraphics[width=\columnwidth]{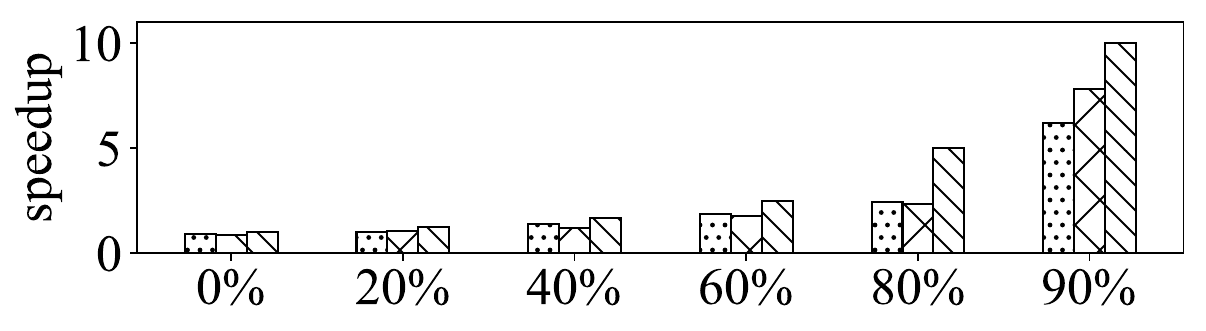}         \vspace{-6mm}
         \caption{PO}
     \end{subfigure}
     \hfill
     \begin{subfigure}[b]{0.49\columnwidth}
         \centering
         \includegraphics[width=\columnwidth]{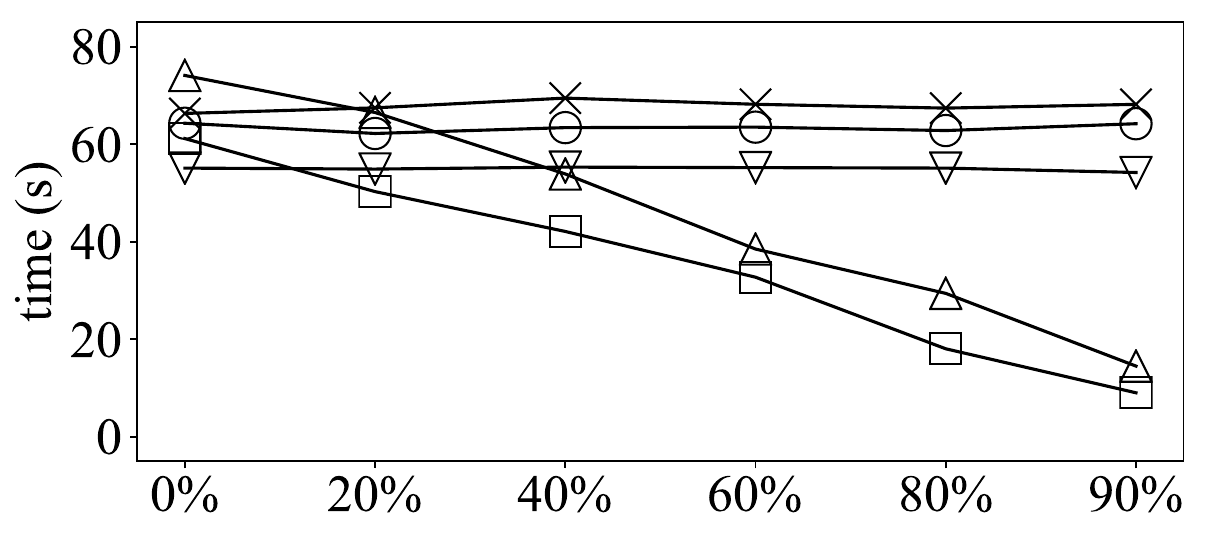}
         \includegraphics[width=\columnwidth]{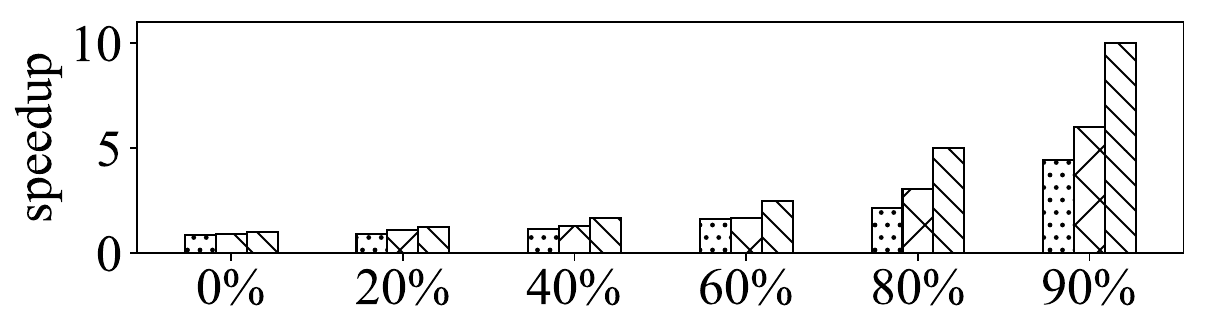}         \vspace{-6mm}
         \caption{LJ}
     \end{subfigure}
          \begin{subfigure}[b]{0.49\columnwidth}
         \centering
         \includegraphics[width=\columnwidth]{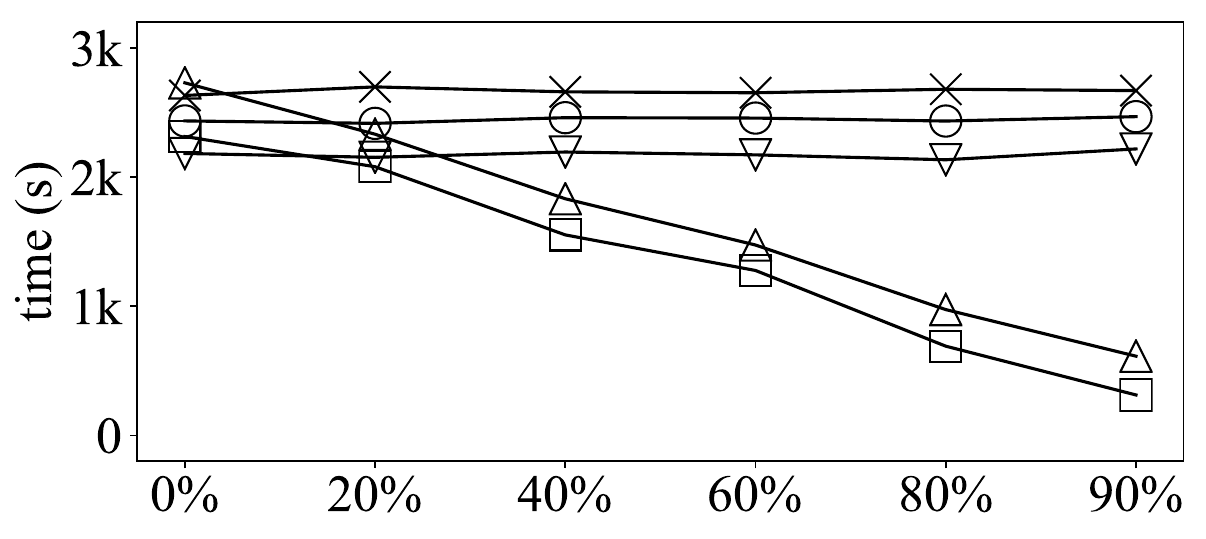}
         \includegraphics[width=\columnwidth]{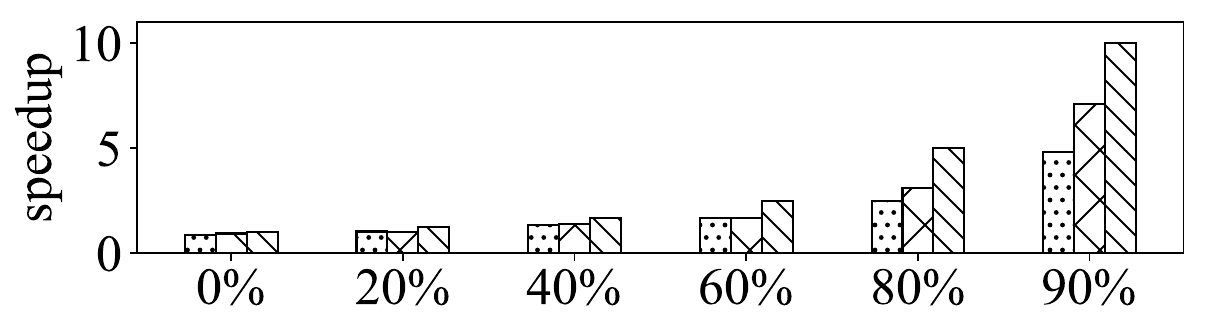}         \vspace{-6mm}
         \caption{TW}
     \end{subfigure}
     \hfill
     \begin{subfigure}[b]{0.49\columnwidth}
         \centering
         \includegraphics[width=\columnwidth]{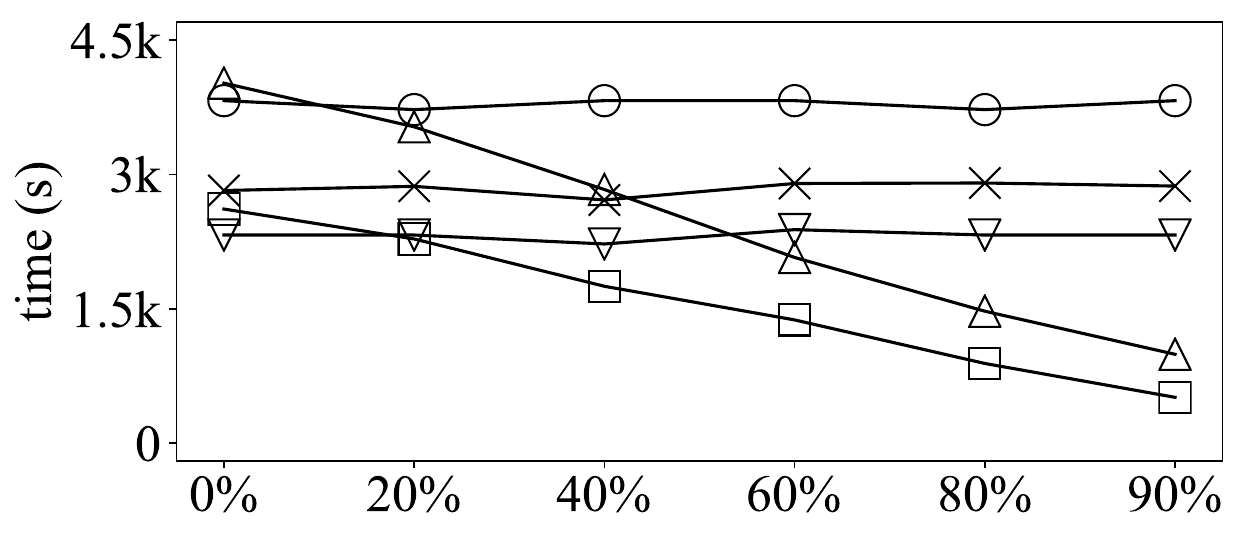}
         \includegraphics[width=\columnwidth]{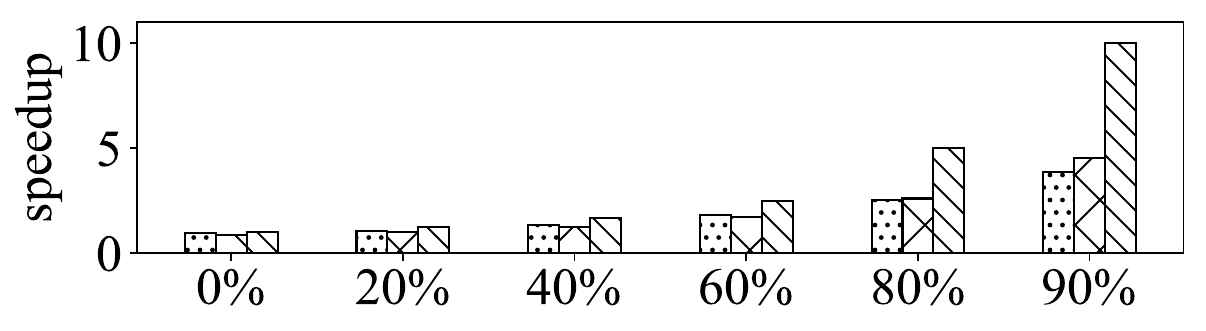}         \vspace{-6mm}
         \caption{FS}
     \end{subfigure}

        \caption{Processing time when varying query similarity}
        \label{fig:vk}
            \vspace{-2mm}

\end{figure*}

\begin{figure*}[!t]
     \centering
      \includegraphics[width=1.2\columnwidth]{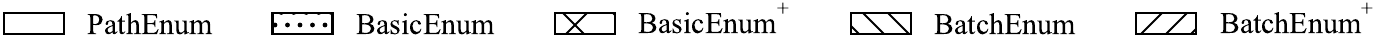}\\

     \begin{subfigure}[b]{0.49\columnwidth}
         \centering
         \includegraphics[width=\columnwidth]{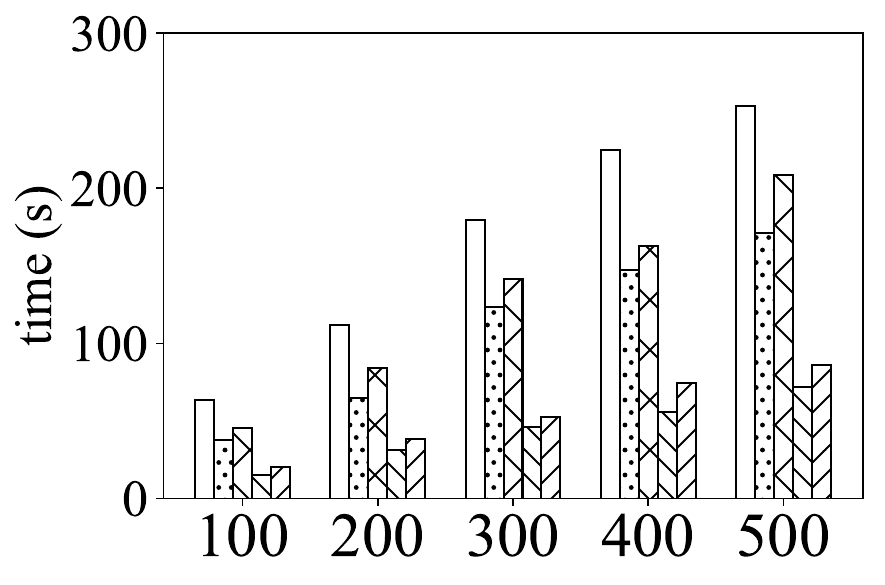}
         \vspace{-6mm}
         \caption{EP}
     \end{subfigure}
     \hfill
     \begin{subfigure}[b]{0.49\columnwidth}
         \centering
         \includegraphics[width=\columnwidth]{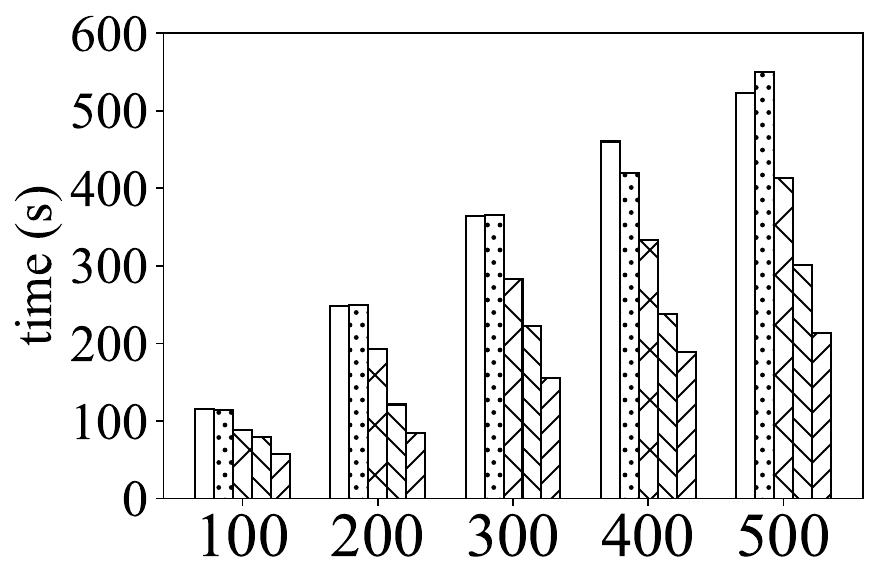}
         \vspace{-6mm}
         \caption{SL}
     \end{subfigure}
          \begin{subfigure}[b]{0.49\columnwidth}
         \centering
         \includegraphics[width=\columnwidth]{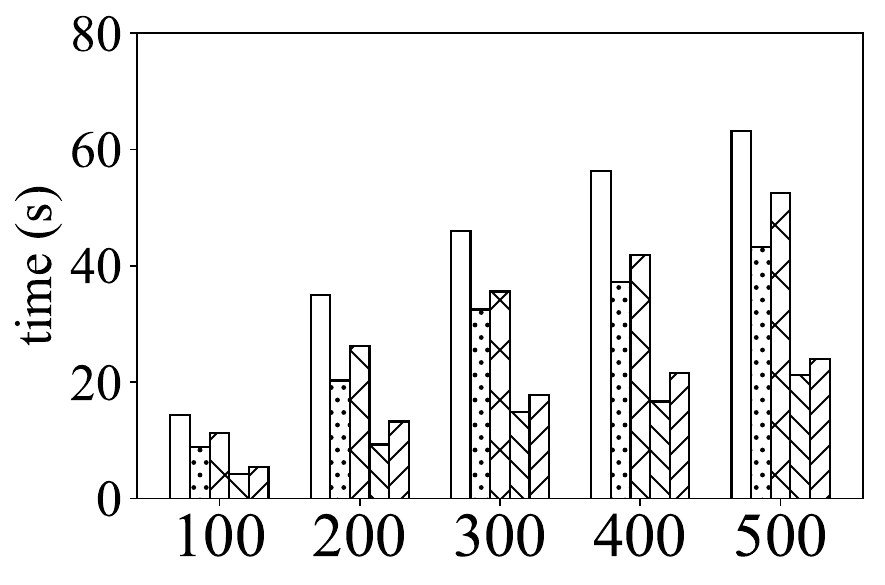}
         \vspace{-6mm}
         \caption{BK}
     \end{subfigure}
     \hfill
     \begin{subfigure}[b]{0.49\columnwidth}
         \centering
         \includegraphics[width=\columnwidth]{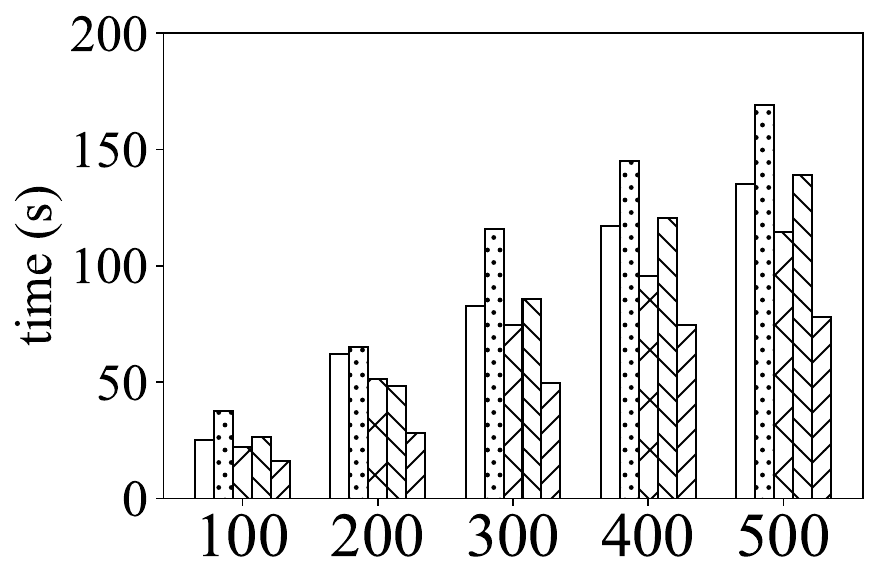}
         \vspace{-6mm}
         \caption{WT}
     \end{subfigure}
          \begin{subfigure}[b]{0.49\columnwidth}
         \centering
         \includegraphics[width=\columnwidth]{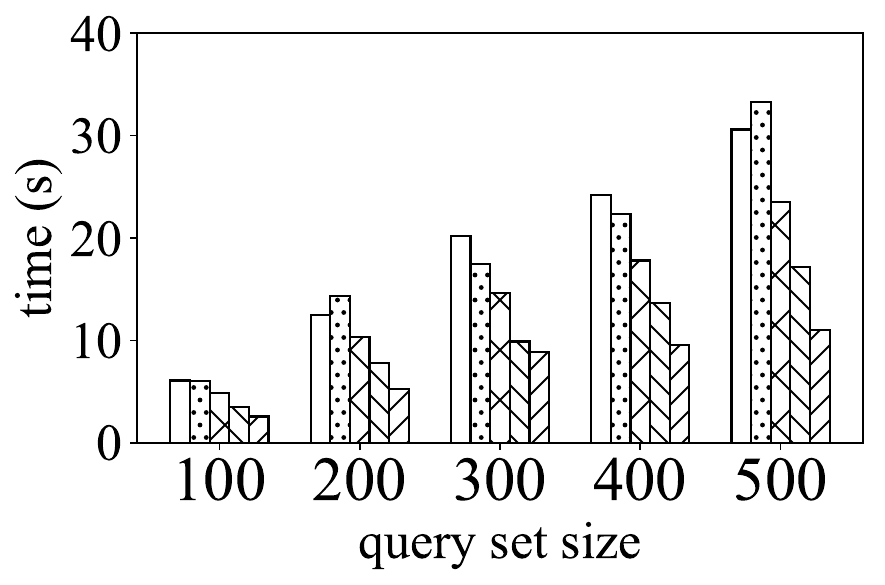}
         \vspace{-6mm}
         \caption{BS}
     \end{subfigure}
     \hfill
     \begin{subfigure}[b]{0.49\columnwidth}
         \centering
         \includegraphics[width=\columnwidth]{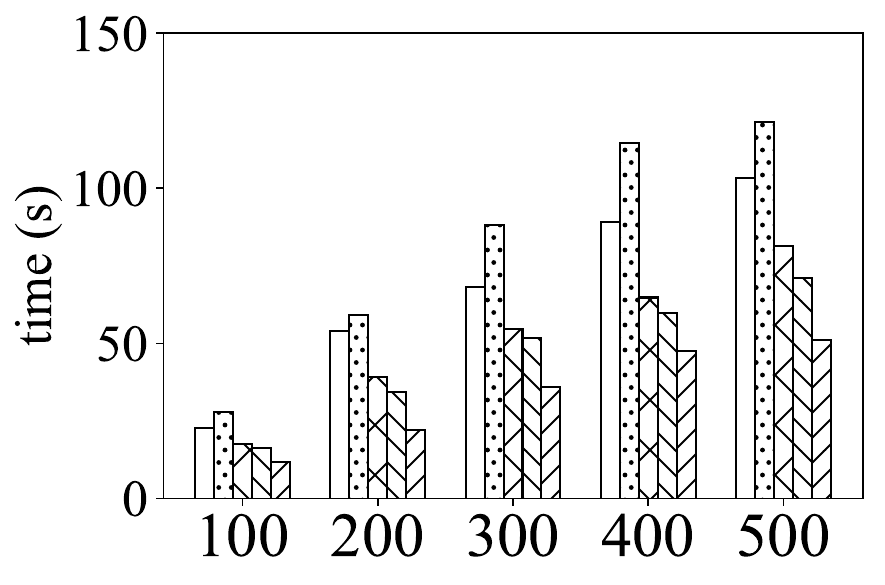}
         \vspace{-6mm}
         \caption{SK}
     \end{subfigure}
          \begin{subfigure}[b]{0.49\columnwidth}
         \centering
         \includegraphics[width=\columnwidth]{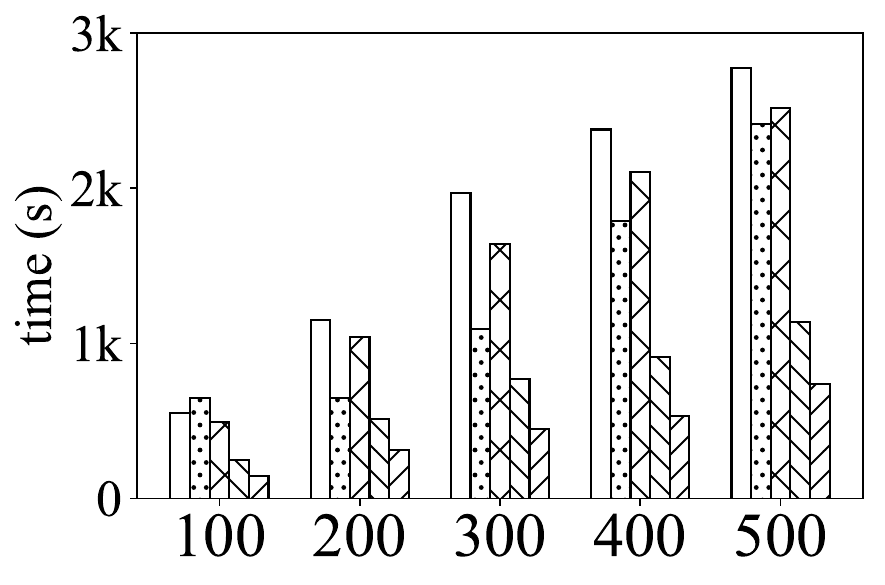}
         \vspace{-6mm}
         \caption{UK}
     \end{subfigure}
     \hfill
     \begin{subfigure}[b]{0.49\columnwidth}
         \centering
         \includegraphics[width=\columnwidth]{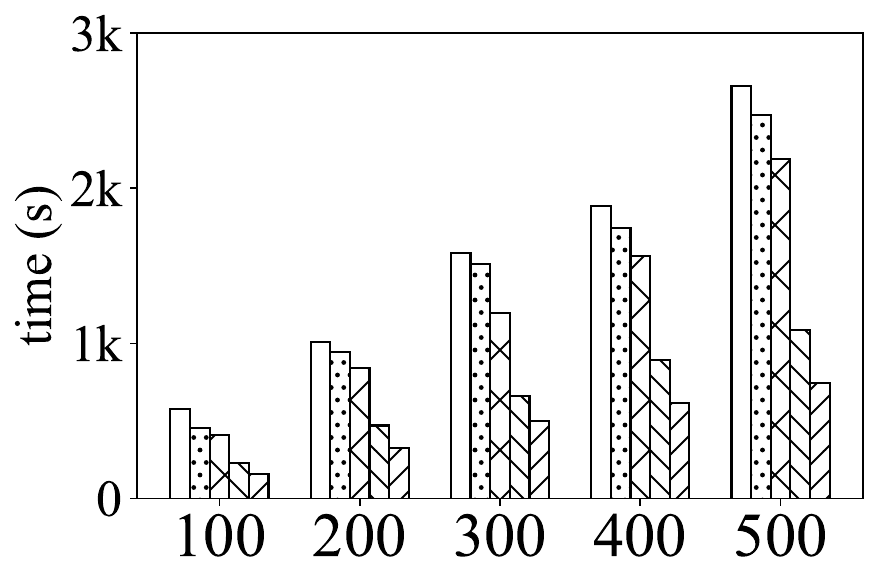}
         \vspace{-6mm}
         \caption{DA}
     \end{subfigure}
    \begin{subfigure}[b]{0.49\columnwidth}
         \centering
         \includegraphics[width=\columnwidth]{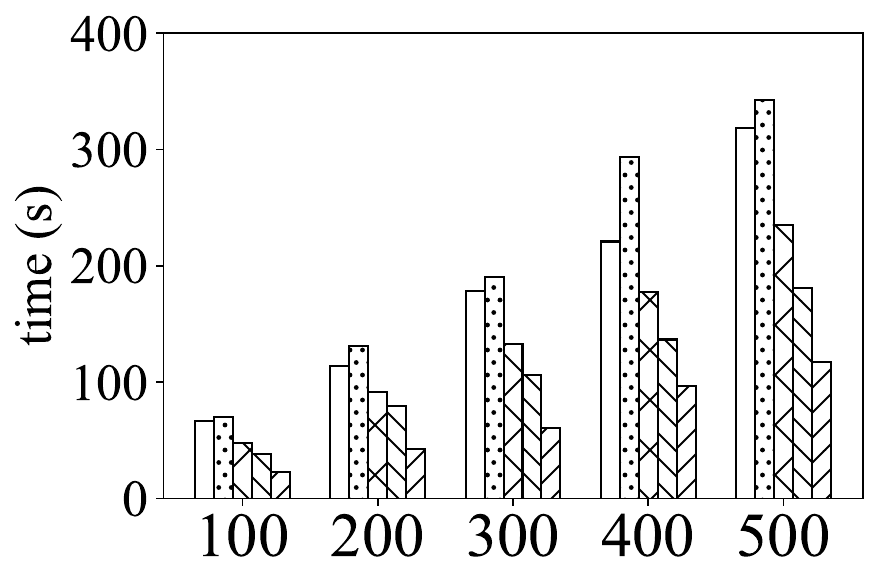}
         \vspace{-6mm}
         \caption{PO}
     \end{subfigure}
     \hfill
     \begin{subfigure}[b]{0.49\columnwidth}
         \centering
         \includegraphics[width=\columnwidth]{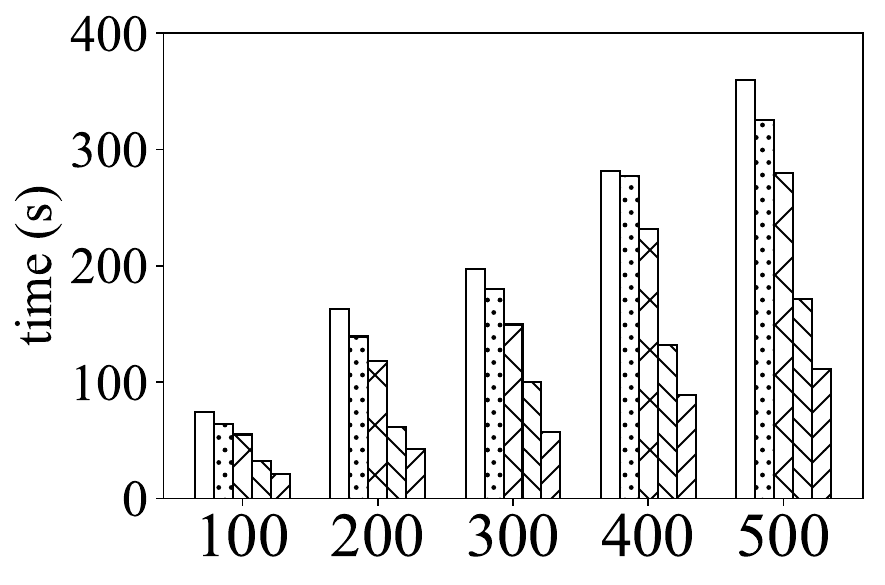}
         \vspace{-6mm}
         \caption{LJ}
     \end{subfigure}
          \begin{subfigure}[b]{0.49\columnwidth}
         \centering
         \includegraphics[width=\columnwidth]{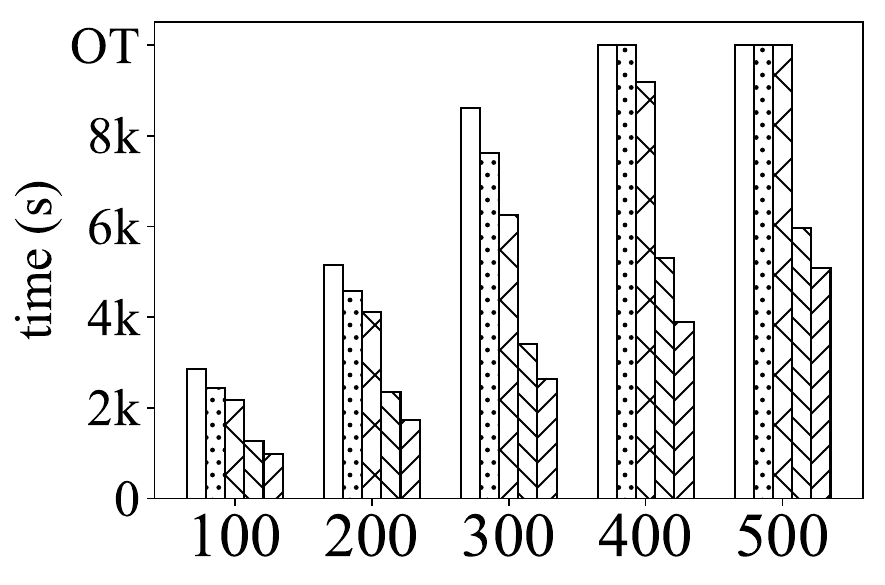}
         \vspace{-6mm}
         \caption{TW}
     \end{subfigure}
     \hfill
     \begin{subfigure}[b]{0.49\columnwidth}
         \centering
         \includegraphics[width=\columnwidth]{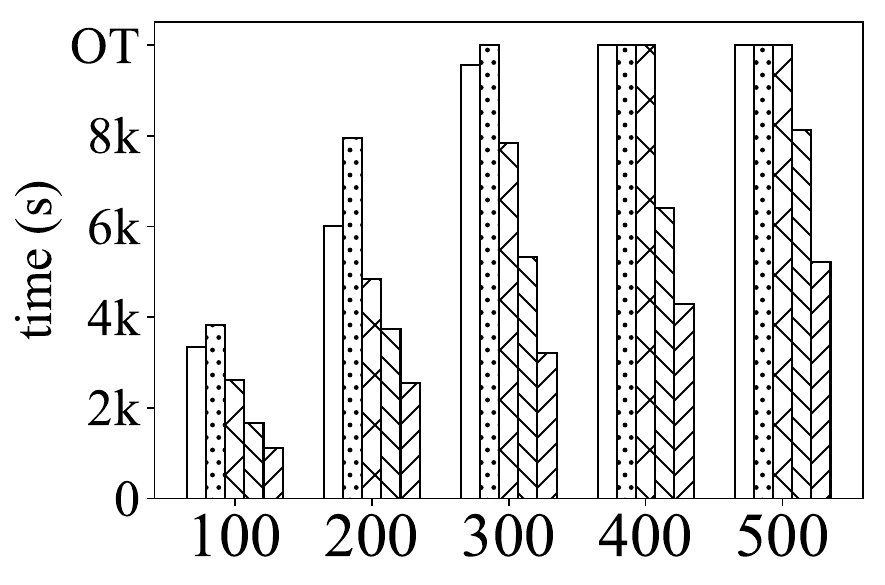}
         \vspace{-6mm}
         \caption{FS}
     \end{subfigure}
                 \vspace{-2mm}

        \caption{Processing time when varying query set size}
        \label{fig:vs}
            \vspace{-2mm}
\end{figure*}

\stitle{Datasets.} We evaluate our algorithms on twelve real-world graphs, which are shown in \reftable{dataset}. Among them,  \kwnospace{Twitter}\textrm{-}\kw{2010} is downloaded from \kw{LAW} (https://law.di.unimi.it/index.php), \kwnospace{Baidu}\textrm{-}\kw{baike}, \kwnospace{Web}\textrm{-}\kwnospace{uk}\textrm{-}\kw{2005} and \kwnospace{Rec}\textrm{-}\kw{dating} are downloaded from \kw{NetworkRepository} (http://networkrepository.com/networks.php), and  the rest are downloaded from \kw{SNAP} (https://snap.stanford.edu/data/). 


\vspace{-1mm}

\stitle{Algorithms.} We compare the following algorithms: 

\begin{itemize}[leftmargin=*]
 \item {\kw{PathEnum}\cite{pathenum}: State-of-the-art \hcstp enumeration algorithm that processes each query independently.}
  \item \pathenum: \refalg{base} in which  \kw{PathEnum} runs bidirectionally (\refsec{exist}).
  \item \pathenump: \refalg{base} in which \kw{PathEnum} runs with an optimized search order (\refsec{exist}).
  \item \batchenum: \refalg{overview} (\refsec{bidirectional}).
  \item \batchenump: \batchenum with an optimized search order introduced by \pathenump (\refsec{exist}).
  \item {\kw{DkSP}: Top-k route planning algorithm proposed in \cite{dksp}.}
  \item {\kw{OnePass}: K-shortest paths algorithm proposed in \cite{onepass}.}
\end{itemize}

\stitle{Settings.} All the algorithms are implemented in \kw{Rust} 1.43. {For \pathenum, \pathenump, \batchenum and \batchenump, we implement their index construction following the state-of-the-art multi-source \kwnospace{BFS}s \cite{vldb14} to reduce the construction time. For \kw{PathEnum}, we directly follow its original implementation to process each query separately. For \kw{DkSP} and \kw{OnePass}, we adapt them to the problem of \hcstp enumeration by ignoring their similarity constraint and keeping generating the path results until reaching the hop constraint.} In the experiments, the time cost is measured as the amount of wall-clock time elapsed during the program's execution. If an algorithm cannot finish in 10,000 seconds, we denote the processing time as \kw{OT}. We set the default query set size as 100 where the hop constraint $k$ of queries uniformly varies from 4 to 7, and the queries are generated by randomly selecting pairs $(s, t)$ such that $s$ can reach $t$ in $k$ hops. We set the default value of $\gamma$ to 0.5.
%

\stitle{Exp-1: Performance when varying query similarity.} In this experiment, we evaluate the effectiveness of our proposed techniques regarding common computation sharing.  Since the precise common computation among queries is hard to obtain, we use the similarity between queries to approximate the common computation and define the similarity of $Q$ as the average query similarity $\mu$ between all pairs of queries in $Q$, i.e. $\mu_Q = \frac{1}{|Q|(|Q|-1)} \cdot \sum_{\forall (q_A,q_B) \in Q \times Q\ s.t.\ q_A \ne q_B}{\mu(q_A, q_B)}$.  We vary the  similarity among the queries in $Q$ from $0\%$ to $90\%$ and report the query processing time and speedup on all datasets. The results are shown in \reffig{vk}. When reporting the speedup, we also show the  speedup limit in which we assume that all common computation is optimally shared and the overhead for common computation sharing is 0s, i.e., the  speedup limit is $1/(1-\mu_Q)$.



As shown in \reffig{vk}, when the query similarity is small such that there is little or no query overlap, both \batchenum and \batchenump can detect it quickly so that the overall processing time is almost the same as sequential processing, with only a slight overhead; on the other hand, when the query similarity grows larger, \batchenum and \batchenump always outperform the other two algorithms with an increasing performance gap. Moreover, it can be seen that when the query set similarity grows larger, the execution times of both \batchenum and \batchenump reduce almost linearly. Remarkably, when the similarity is set to 90\%, the speedup limit is $1/(1-0.9)=10$ times, while our algorithm can achieve a maximum speedup of 7.7 times and an average speedup of 6.4 times. This verifies the effectiveness of our computation sharing mechanism in \batchenum and \batchenump. For \pathenum and \pathenump, \pathenump is faster in most cases, which is consistent with the analysis in \cite{pathenum}. {Additionally, \pathenump is always faster than \kw{PathEnum} due to the common computation sharing across queries.} 


\vspace{-0.5mm}

\stitle{Exp-2: Efficiency when varying $|Q|$.} In this experiment, we evaluate the efficiency of the algorithms when varying the query set size $|Q|$. We randomly generate query sets with size from 100 to 500 and report the processing time for each query set  in \reffig{vs}. As can be seen, \batchenum and \batchenump outperform the three baseline algorithms on all datasets. Remarkably, for the two billion-scale graphs \kw{TW} and \kw{FS}, the three baseline algorithms fail on both of these graphs when the query set size grows larger; comparatively, our algorithms always efficiently finish on these graphs for all the query sets. {Based on the results, it is clear that \batchenum and \batchenump are considerably more efficient than the compared algorithms. }

\begin{figure}
         \centering
         \includegraphics[width=\columnwidth]{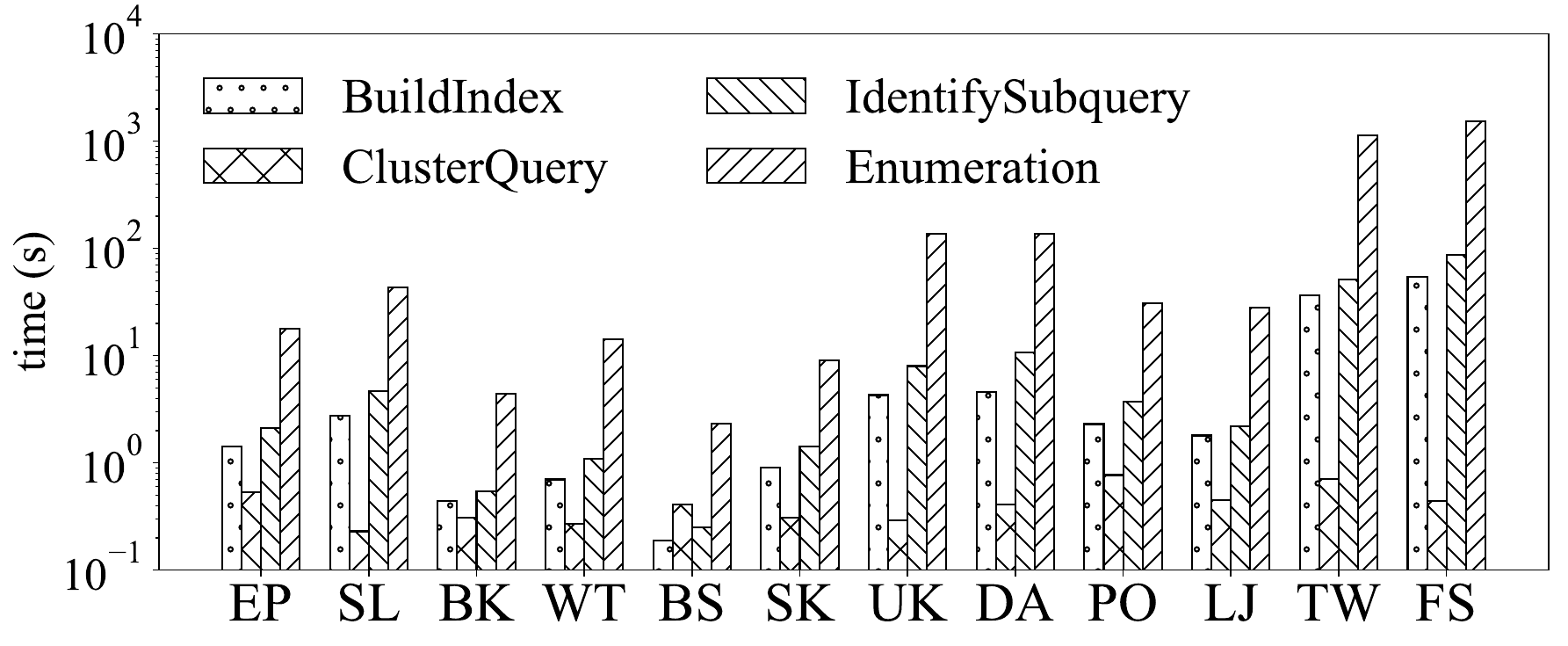}
         \vspace{-7mm}
         \caption{Processing time decomposition}
        \label{fig:td}

\end{figure}

\begin{figure}
         \centering
         \includegraphics[width=0.8\columnwidth]{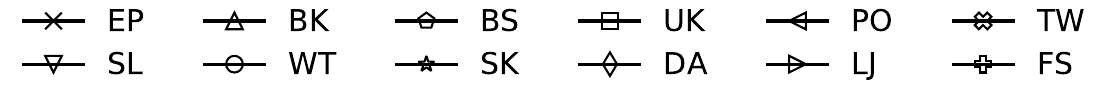}
         \includegraphics[width=0.9\columnwidth]{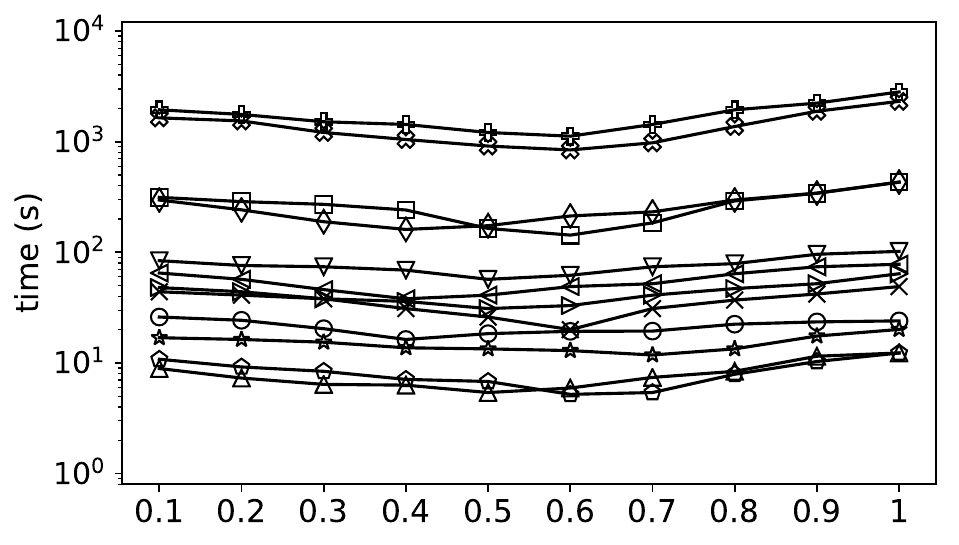}
         \vspace{-3mm}
         \caption{Impact of $\gamma$}
        \label{fig:tg}
\end{figure}

\begin{figure}
     \centering
           \includegraphics[width=\columnwidth]{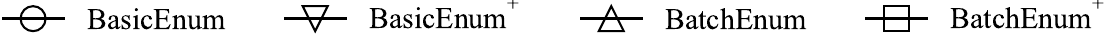}\\
              \begin{subfigure}[b]{0.49\columnwidth}
         \centering
         \includegraphics[width=\columnwidth]{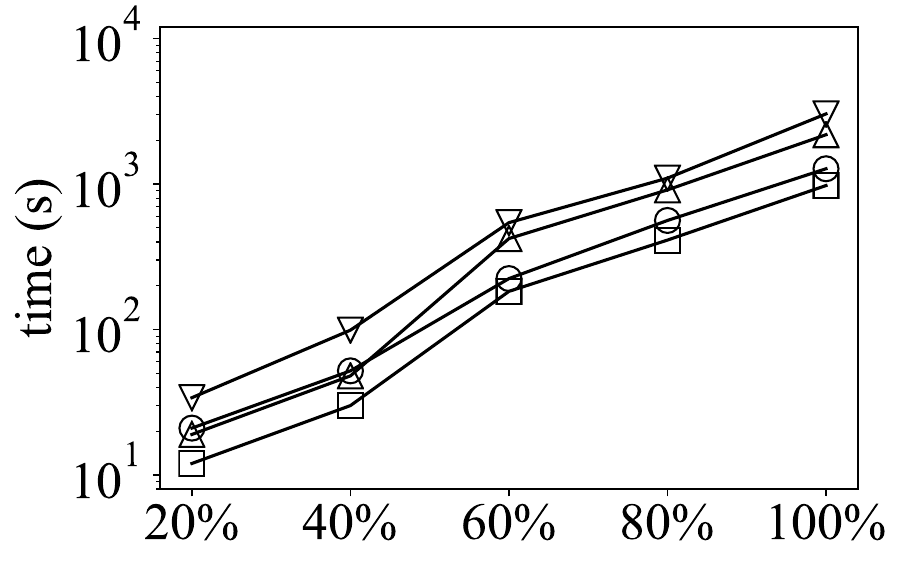}
         \vspace{-2mm}
         \caption{TW (Vary $|V(G)|$)}
     \end{subfigure}
     \begin{subfigure}[b]{0.49\columnwidth}
         \centering
         \includegraphics[width=\columnwidth]{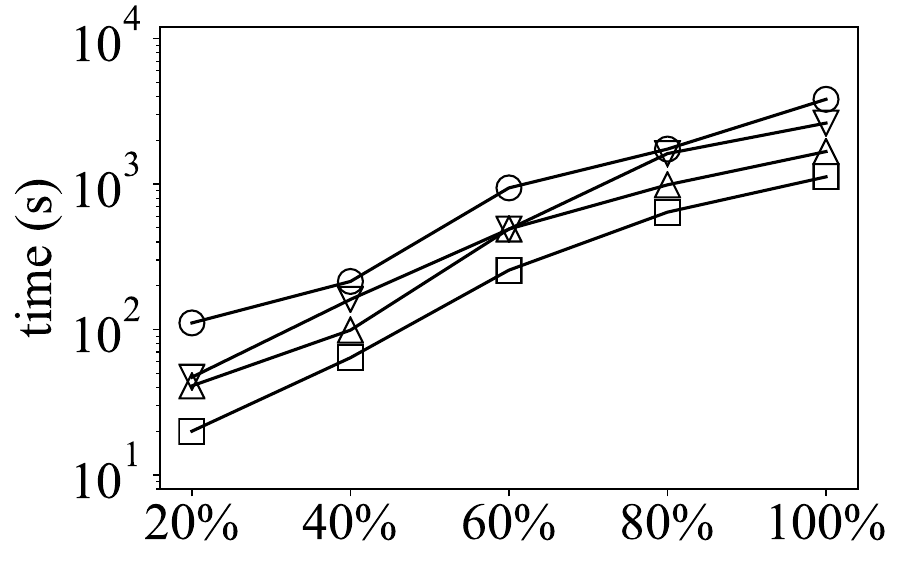}
         \vspace{-2mm}
         \caption{FS (Vary $|V(G)|$)}
     \end{subfigure}

  \caption{Processing time when varying graph size}
  \label{fig:sample}
      \vspace{-1mm}
\end{figure}


\stitle{Exp-3: Processing time decomposition.} In this experiment, we report the  time consumed in each sub-step of \batchenump, including \kw{BuildIndex}, \kw{ClusterQuery}, \kw{IdentifySubquery}, and \kw{Enumeration}. Specifically, the \kw{BuildIndex} time is the time spent on building the index for multiple queries, the \kw{ClusterQuery} time is that spent on clustering queries into groups (\refalg{cluster}), the \kw{IdentifySubquery} time is that spent on detecting the common \hcsp queries in the groups (\refalg{identify}) and the \kw{Enumeration} time is that spent on path enumeration after finding the \hcsp queries. As shown in \reffig{td}, the processing time of \batchenump is dominated by the \kw{Enumeration} time on all graphs, which verifies that the impact of computation sharing overhead is limited. {Moreover, it can be seen that the \kw{ClusterQuery} time is small on all graphs. This is because the cost of \querycluster is  dependent on $|Q|$ and the hop-constrained neighbors of $q \in Q$, which are generally much smaller than the graph size.} Additionally, \kw{BuildIndex} and \kw{IdentifySubquery} only perform BFS traversals on the graph, hence their time is much smaller than the \kw{Enumeration} time. 

\stitle{Exp-4: Efficiency regarding $\gamma$.} We evaluate the impact of threshold $\gamma$ regarding the query clustering on our algorithm. We vary the value of $\gamma$ from 0.1 to 1 and report the processing time of our algorithm on all datasets in \reffig{tg}. \reffig{tg} shows that: (1) as $\gamma$ decreases, the processing time decreases until $\gamma$ reaches a turning point. (2) when $\gamma$ keeps decreasing after reaching the turning point, the processing time increases again.  This is because as $\gamma$ decreases, the number of queries in the same sharing group will increase, which  provides more opportunities for computation sharing. When too many queries, with smaller similarity, are allocated to the same group, many of them may not necessarily have common \hcsp queries that are worth extracting. Thus, the processing time increases due to the overhead. 

%
%
%
%
%

\stitle{Exp-5: Scalability.} In this experiment, we evaluate the scalability of the algorithms as the graph grows in size. We choose the two largest datasets \kw{TW} and \kw{FS} and randomly sample their vertices and edges from 20\% to 100\%. The processing time is shown in \reffig{sample} (Varying number of edges is not shown due to the similar trends). As can be seen, when the graph size increases, the processing time of all the algorithms also increases. This is because as the graph size increases, the search space rapidly grows larger. Moreover, it can be seen that \batchenum and \batchenump outperform the other baseline algorithms in all cases. This is attributed to our computation sharing strategy, which effectively reduces the overall computation cost and therefore, increases the algorithm's scalability.

\begin{figure}
         \vspace{-5mm}
         \centering
         \includegraphics[width=0.7\columnwidth]{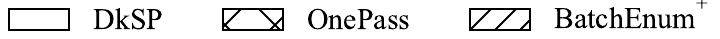}
         \includegraphics[width=\columnwidth]{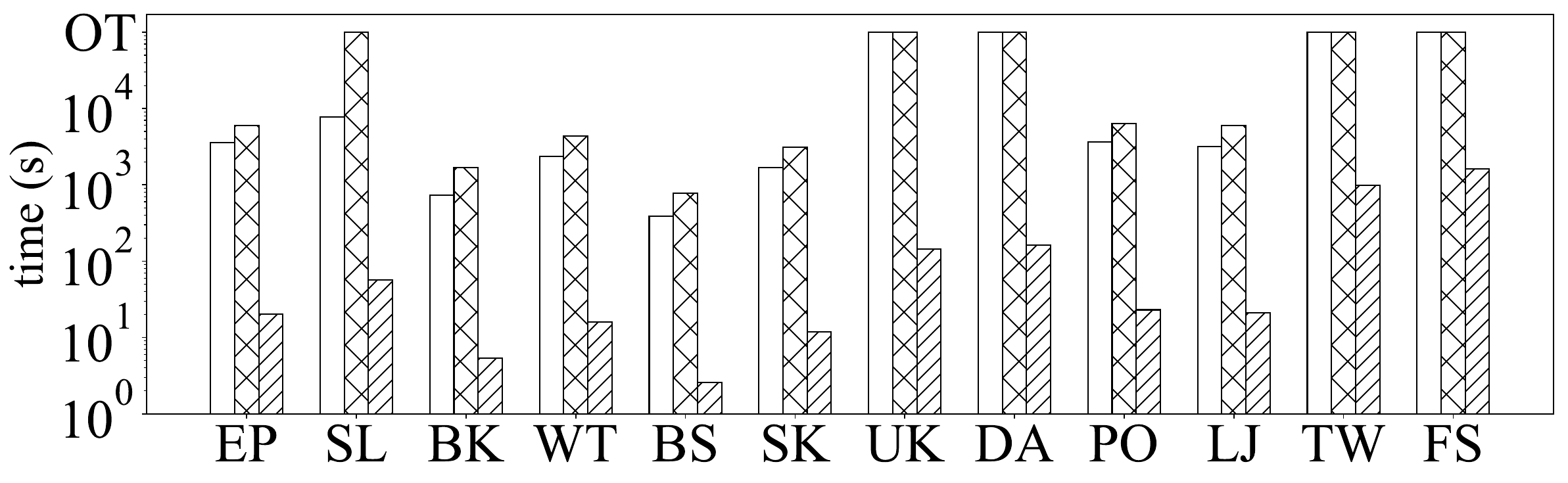}
         \vspace{-5mm}
         \caption{{Processing time compared with KSP algorithms}}
        \label{fig:new_cmp}
\end{figure}

\stitle{{Exp-6: Efficiency comparison with $k$ shortest path algorithms.}} {In this experiment, we compare the efficiency of our algorithm with the adapted $k$ shortest paths algorithms \kw{DkSP} and \kw{OnePass}. We evaluate their performance on all datasets by randomly generating 100 queries with hop constraint $k$ varying from 3 to 7. As can be seen in \reffig{new_cmp}, our approach outperforms the $k$ shortest paths algorithms by over two orders of magnitude. This is because \kw{DkSP} and \kw{OnePass} do not consider the pruning opportunities specific to \hcstp enumeration, which is crucial as proven in \cite{pathenum}.}

\begin{figure}
         \centering
         \includegraphics[width=0.8\columnwidth]{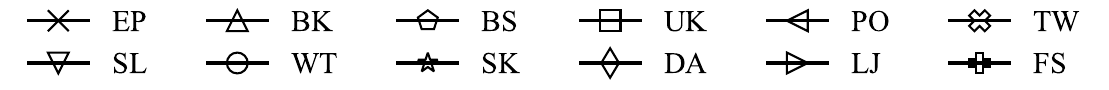}
         \includegraphics[width=0.8\columnwidth]{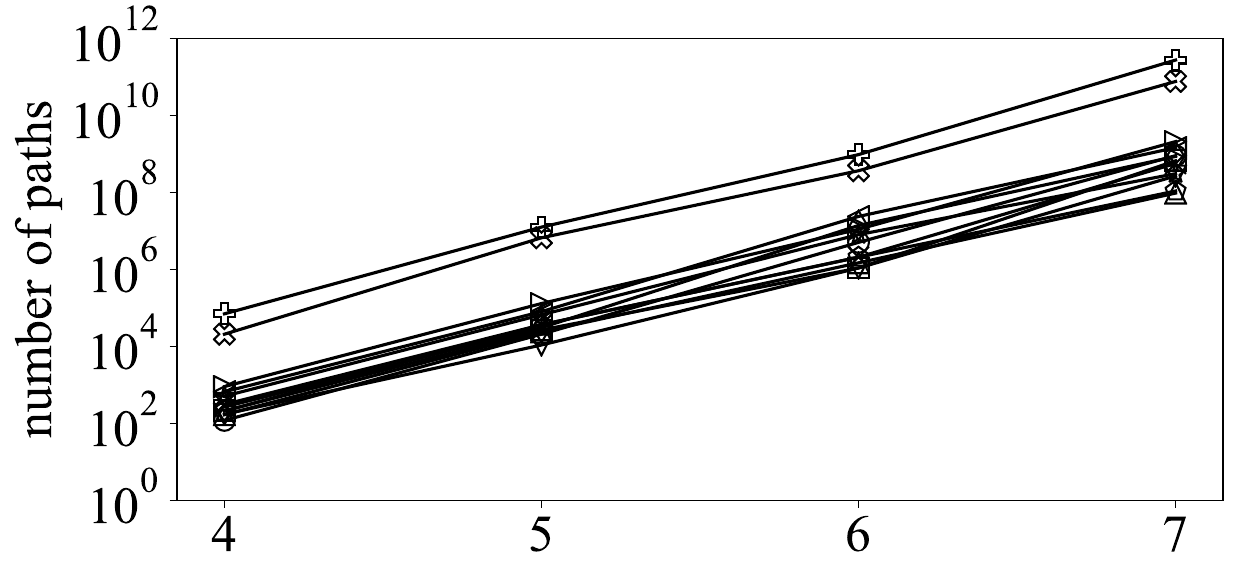}
         \vspace{-3mm}
         \caption{{Average number of paths when varying $k$}}
        \label{fig:path_number}
\end{figure}

\stitle{{Exp-7: Number of paths for \hcstp query.}} {In this experiment, we randomly generate 100 queries with the specific hop constraint  $k$ from 3 to 7, and report the average of number \hcstps for each query on all datasets. As shown in \reffig{path_number}, the average number of \hcstps for each query grows exponentially with $k$.}

\vspace{-1mm}
\section{Related Works}
\label{sec:related} 




\stitle{HC-s-t path enumeration.} Research efforts have been devoted to many fundamental problem in graph analysis \cite{liu2020efficient,chen2023index,zhang2022shortest,gao2020time,wang2022promptem}. \hcstp enumeration is a fundamental problem in graph analysis and several algorithms  for this problem have been proposed  \cite{Ri14, Gr18, Qi18, Pe19, pathenum}. Generally, the algorithms can be divided into two categories: pruning-based algorithms \cite{Ri14, Gr18, Pe19} and index-based algorithms \cite{pathenum}. (1) Pruning-based algorithms. Pruning-based algorithms typically adopt a backtracking strategy based on a depth-first search based framework. During the enumeration, \cite{Ri14} and \cite{Gr18} dynamically compute the shortest path distance from $v$ to $t$ and prune $v$ if it is unreachable to $t$, while \cite{Pe19} dynamically maintains a lower bound of hops to the target vertex $t$ for the vertices visited and prunes $v$ if the current remaining hop budget is smaller than the lower bound of hops required. (2) Index-based algorithm. Sun et al. \cite{pathenum} finds that the pruning-based algorithms typically suffer from severe performance issues caused by the costly pruning operations during enumeration. Therefore, \kw{PathEnum} builds a light-weight index to reduce the number of edges involved in the enumeration, which has been clearly presented in \refsec{exist}. However, all of the above algorithms focus on improving the performance of a single \hcstp query and do not consider common computation sharing for batch processing.

\stitle{Batch query processing for other graph problems.} There are a number of works on batch  query processing for other graph problems \cite{Re16, lumulti, Am21, Le12,thomsen2012effective,thomsen2014concise,li2020fast}.  Le et al. \cite{Le12} studies the batch processing of \kw{SPARQL} queries on RDF graphs.  Ren el al. \cite{Re16} studies the batch processing of subgraph matching queries, which extracts and pre-computes the query graphs' overlaps by finding their maximum common subgraphs. Mhedhbi et al. \cite{Am21} also studies the batch processing of subgraph matching queries, which uses a greedy cost-based optimizer to compute a combined join plan for all subgraph queries.  Thomsen et al. \cite{thomsen2012effective} uses a cache to store the most beneficial paths such that all the sub-path queries of the cached paths can be answered directly. Li et al. \cite{li2020fast} proposes three query decomposition methods to cluster queries and two batch algorithms to re-use the cached results. 

\section{Conclusion}
\label{sec:conclusion} 

{In this paper, we study the batch processing of \hcstp queries. We first show that finding an optimal solution for batch HC-s-t path query processing is NP-hard. Therefore, we aim to design a practically efficient algorithm to accelerate the batch HC-s-t path query processing. To achieve this goal, we observe that there exists common computation among multiple queries and the processing performance can be significantly improved if these common computation can be effectively shared. Following this observation, we first propose a two-phase algorithm to detect the common sub-structures among queries. Based on the identified common sub-structures, we design an efficient algorithm to process the given queries by sharing the computation related to  the identified common sub-structures. Due to re-using the common computation effectively, the experimental results  demonstrate the efficiency and scalability of our proposed algorithms.}

\balance
\bibliography{sample}
\bibliographystyle{IEEEtran}

\end{document}